# Energy-Aware Competitive Power Allocation for Heterogeneous Networks Under QoS Constraints

Giacomo Bacci, *Member, IEEE*,  E. Veronica Belmega, *Member, IEEE*,
Panayotis Mertikopoulos, *Member, IEEE*,  and Luca Sanguinetti, *Member, IEEE*

## Abstract

This work proposes a distributed power allocation scheme for maximizing energy efficiency in the uplink of OFDMA-based HetNets where a macro-tier is augmented with small cell access points. Each user equipment (UE) in the network is modeled as a rational agent that engages in a non-cooperative game and allocates its available transmit power over the set of assigned subcarriers to maximize its individual utility (defined as the user's throughput per Watt of transmit power) subject to a target rate requirement. In this framework, the relevant solution concept is that of Debreu equilibrium, a generalization of the concept of Nash equilibrium. Using techniques from fractional programming, we provide a characterization of equilibrial power allocation profiles. In particular, Debreu equilibria are found to be the fixed points of a water-filling best response operator whose water level is a function of rate constraints and circuit power. Moreover, we also describe a set of sufficient conditions for the existence and uniqueness of Debreu equilibria exploiting the contraction properties of the best response operator. This analysis provides the necessary tools to derive a power allocation scheme that steers the network to equilibrium in an iterative and distributed manner without the need for any centralized processing. Numerical simulations are used to validate the analysis and assess the performance of the proposed algorithm as a function of the system parameters.

G. Bacci is with Mediterranean Broadband Infrastructure (MBI) srl, 56121 Pisa, Italy (email: gbacci@mbigroup.it).

E. V. Belmega is with ETIS/ENSEA, Université de Cergy-Pontoise, 95014 Cergy-Pontoise, France (email: elena-veronica.belmega@ensea.fr).

P. Mertikopoulos is with the French National Center for Scientific Research (CNRS) and the Laboratoire d'Informatique de Grenoble, Grenoble, France (email: panayotis.mertikopoulos@imag.fr).

L. Sanguinetti is with the Dipartimento di Ingegneria dell'Informazione, University of Pisa, 56122 Pisa, Italy (email: luca.sanguinetti@unipi.it), and also with the Large Networks and Systems Group, CentraleSupélec, 91192 Gif-sur-Yvette, France.

## I. INTRODUCTION

Owing to the prolific spread of Internet-enabled mobile devices and the ever-growing volume of mobile communication calls, the biggest challenge in the wireless industry today is to meet the soaring demand for wireless broadband required to ensure consistent quality of service (QoS) in a network. Rising to this challenge means increasing the network capacity by a thousandfold over the next few years [1], but the resulting power consumption and energy-related pollution are expected to give rise to major societal, economic and environmental issues that would render this growth unsustainable [2]. Therefore, the information and communications technology (ICT) industry is faced with a formidable mission: cellular network capacity must be increased significantly in order to accommodate higher data rates, but this task must be accomplished under an extremely tight energy budget.

A promising way out of this gridlock is the small-cell (SC) network paradigm which builds on the premise of shrinking wireless cell sizes in order to bring user equipment (UE) and their serving stations closer to one another. From an operational standpoint, SC networks can be integrated seamlessly into existing macro-cellular networks: the latter ensure wide-area coverage and mobility support, while the former carry most of the generated data traffic [3].

Albeit promising, the deployment of this kind of networks, commonly referred to as heterogeneous networks (HetNets), poses several technical challenges mainly because different SCs are likely to be connected over unreliable infrastructures with widely varying features – such as error rate, outage, delay, and/or capacity specifications. Accordingly, the inherently heterogeneous nature of these networks calls for flexible and decentralized resource allocation strategies that rely only on local channel state information (CSI) and require minimal information exchange between network users and/or access points/base stations. This framework is commonly referred to as *distributed optimization*, and it represents a crucial aspect of scalable and efficient network operation.

An established theoretical tool for problems of this kind is provided by the theory of *non-cooperative games* [4]. Among the early contributions in this area, [5, 6] investigated the rate maximization problem for autonomous digital subscriber lines based on competitive optimality criteria. In the spirit of these works, a vast corpus of literature has since focused on developing power control techniques for unilateral spectral efficiency maximization subject to individual power constraints. For instance, [7, 8] proposed a game-theoretic approach to energy-efficient power control in multi-carrier code division multiple access (CDMA) systems, [9–12] investigated the problem of distributed power control in multi-user multiple-input and multiple-output (MIMO) systems, [13, 14] studied the interference relay channel, while two-tier CDMA networks were examined in [15]. More recently, the authors of [16] used a variational inequality (VI) framework to model and analyze the competitive spectral efficiency maximization problem. The analogy between Nash equilibria and VIs was subsequently exploited in [17] to design distributed power control algorithms for spectral efficiency maximization under interference temperature constraints in a cognitive radio context.

Distributed power allocation policies as above have the important advantage of avoiding the waste of energy associated with centralized algorithms requiring considerable information exchange (and, hence, transmissions) between the users and/or the network administrator [16]. On the other hand, the users' aggressive attitude towards interference from other users can lead to a cascade of power increases at the UE level, thereby leading to battery depletion and inefficient energy use. Consequently, solutions that focus exclusively on spectral efficiency maximization are not aligned with energy-efficiency requirements [18, 19] – which, as we mentioned above, are crucial for the deployment and operation of HetNets.

### A. Summary of contributions

Our main goal in this paper is the analysis and design of energy-efficient power allocation policies in a HetNet setting where SC networks coexist with macro-tier cellular systems based on orthogonal frequency-division multiple access (OFDMA) technology. In particular, focusing on the uplink case, we propose a game-theoretic framework where each UE adjusts the allocation of its transmit power (over the available subcarriers) so as to unilaterally maximize its individual link utility subject to a minimum rate requirement. Specifically, each user's energy-aware utility function is defined as the achieved throughput per unit power, accounting for both the power required for data transmission and that required by the circuit components of each UE (such as amplifiers, mixer, oscillator, and filters) [20–22].

Due to each user's rate constraints, the resulting game departs from the classical framework put forth by Nash [23] and gives rise to a Debreu-type game [24] where the actions available to each UE depend on the transmit power profile of all other users in the network. In this setting, the relevant solution concept is that of a *Debreu equilibrium* (DE) [24] – also known as a generalized Nash equilibrium (GNE) [25]. Drawing on fractional programming techniques [26], we characterize the system's Debreu equilibria as fixed points of a water-filling operator whose water level is a function of the users' minimum rate constraints and circuit power [22]. This characterization is then used to provide sufficient conditions for DE uniqueness and to derive a distributed power allocation algorithm that allows the network to converge to equilibrium under minimal information assumptions. The performance of the proposed solution is then validated by means of extensive numerical simulations modeling a HetNet where a macro-tier is augmented with a certain number of low range small-cell access points (SCAs). As it turns out, the proposed solution represents a scalable and flexible technique to meet the ambitious goals of 5G communications [27], such as extremely high area spectral efficiency (ASE) (more than $500 \, \text{b/s/Hz/km}^2$) with a reasonable amount of physical resources (bandwidth and power) and complexity at the network level (number of SCs, signal processing burden, and number of transmit and receive antennas).

Our work builds on the game-theoretic analysis proposed in [28] where a group of players aims at maximizing their individual energy efficiency (EE) (measured in bits per Watt of transmit power) subject to each user's power constraints. Despite this similarity, the analysis of [28] does not account for minimum rate requirements, thus the resulting game-theoretic model is a standard Nash game with no QoS guarantees – in particular, the users' rates at equilibrium could be fairly low. Incorporating QoS requirements changes the setting drastically and takes us beyond the standard Nash framework because a user's admissible power allocation policy depends crucially on the transmit powers of all other users. The energy-efficient framework proposed in this paper represents a generalization of the power minimization under minimum-rate constraints investigated in [29], which is a special case that occurs when the minimum rates are achieved with equality. Preliminary versions of our results appeared in the conference paper [30]: in contrast to this earlier paper, we provide here a complete equilibrium analysis and characterization along with sufficient conditions that guarantee the convergence of the system to a stable equilibrium state.

*B. Paper outline and notation*

The remainder of this paper is organized as follows. In Section II, we introduce the system model and the EE maximization problem with minimum rate constraints. In Section III, we first formulate the non-cooperative game and then study the existence and uniqueness of Debreu equilibria. Section IV presents an iterative and distributed algorithm to reach the equilibrium point, whereas Section V reports numerical results that are used to assess the performance of the proposed solution and to make comparisons with alternatives. Conclusions and perspectives are presented in Section VI.

Matrices and vectors are denoted by bold letters, $\mathbf{I}_L$, $\mathbf{0}_L$, and $\mathbf{1}_L$ are the $L \times L$ identity matrix, the $L \times 1$ all-zero column vector, and the $L \times 1$ all-one column vector, respectively, and $\|\cdot\|$, $(\cdot)^T$ and $(\cdot)^H$ denote Euclidean norm of the enclosed vector, transposition and Hermitian conjugation respectively. The notation $(x)^+$ stands for $\max\{0, x\}$ whereas $W(\cdot)$ denotes the Lambert $W$ function [31], defined as the multiple-branch solution of the equation $z = W(z)\, e^{W(z)}$, $z \in \mathbb{C}$. $\mathbb{1}_X$ denotes the indicator function such that $\mathbb{1}_X = 1$ if $X$ is true, and $0$ elsewhere. Finally, if $\mathcal{A}_k$, $k = 1, \ldots, K$, is a finite family of sets, and $a_k \in \mathcal{A}_k$, we will use the notation $(a_k; \mathbf{a}_{-k}) \in \prod_k \mathcal{A}_k$ as shorthand for the profile $(a_1, \ldots, a_k, \ldots, a_K)$, and $|\mathcal{A}_k|$ to denote its cardinality.

## II. System Model and Problem Formulation

*A. System model*

We consider the uplink of a slowly-varying HetNet where $S$ low-range SCAs are adjoined to a macro-tier cell operating in an OFDMA-based open-access licensed spectrum. For notational compactness, we will reserve the index $s = 0$ for the macrocell base station (MBS), so that $\mathcal{S} = \{0, 1, \ldots, S\}$ represents the set of HetNet receiving stations. The $s$-th cell uses a set of orthogonal subcarriers to serve the $K_s$ user equipment (UE) falling within its coverage radius $\rho_s$. For simplicity, we assume that the same set of subcarriers $\mathcal{N} = \{1, \ldots, N\}$ is used by both tiers. We also assume that $\mathcal{N}$ is assigned by the network and cannot be controlled by the cell operators. Each cell access point (AP) is further equipped with $M_s$ receiving antennas, whereas a single antenna is employed at the UE to keep the complexity of the front-end limited. The framework described in the paper can be generalized to the case of a multicellular HetNet scenario (including MIMO configurations) in a straightforward manner.

Let $\mathbf{h}_{kj,n} \in \mathbb{C}^{M_{\psi(k)} \times 1}$ denote the uplink channel vector with entries $[\mathbf{h}_{kj,n}]_m$ representing the (frequency) channel gains over subcarrier $n$ from the $j$-th UE to the $m$-th receive antenna of the serving AP $\psi(k)$ of user $k$, where $\psi(k) : \mathcal{K} \mapsto \mathcal{S}$ is a generic function that assigns each user $k$ its serving AP.[1] In the following, $\mathcal{K} = \{1, \ldots, K\}$ and $K = \sum_{s=0}^{S} K_s$ denote the set and the number of UE in the network respectively, with $K_s$ representing the number of UE in the $s$-th cell: if $s = 0$, the UE will be termed macrocell user equipment (MUE), and small-cell user equipment (SUE) otherwise, although there is no substantial distinction among the two classes of users (this is clarified further in the rest of this paper). We also assume that the channels remain constant within a reasonable time interval (for more quantitative details, see Section V).

We let $z_{j,n}$ denote the data symbol of UE $j$ over subcarrier $n$ and write $p_{j,n}$ for its corresponding power. The vector $\mathbf{x}_{k,n} \in \mathbb{C}^{M_{\psi(k)} \times 1}$ collecting the samples received over subcarrier $n$ at the AP serving the $k$-th UE can then be written as

$$\mathbf{x}_{k,n} = \sqrt{p_{k,n}}\mathbf{h}_{kk,n}z_{k,n} + I_{k,n} + \mathbf{w}_{k,n} \tag{1}$$

where $\mathbf{w}_{k,n} \sim \mathcal{CN}(\mathbf{0}_{M_{\psi(k)}}, \sigma^2 \mathbf{I}_{M_{\psi(k)}})$ is thermal noise and

$$I_{k,n} = \sum_{j=1, j \neq k}^{K} \sqrt{p_{j,n}}\mathbf{h}_{kj,n}z_{j,n} \tag{2}$$

accounts for the multiple access interference (MAI) experienced by user $k$ over subcarrier $n$. Note that $I_{k,n}$ accounts for both intra-cell interference (generated by other UE served by the same AP) and inter-cell interference (from UE served by all other APs). To keep the complexity at a tolerable level, a simple linear detection scheme is employed for data detection, although a generalization to nonlinear detectors is straightforward. This means that the entries of $\mathbf{x}_{k,n}$ are linearly combined to form $y_{k,n} = \mathbf{g}_{k,n}^H \mathbf{x}_{k,n}$ where $\mathbf{g}_{k,n}$ is the vector employed for recovering the data transmitted by user $k$ over subcarrier $n$. Then, the signal-to-interference-plus-noise ratio (SINR) over the $n$-th subcarrier that is achieved by user $k$ at its serving AP takes the form:

$$\gamma_{k,n} = \mu_{k,n}(\mathbf{p}_{-k,n})p_{k,n} \tag{3}$$

---

[1] For a more detailed description of this assignment mapping, see Section V.

where $\mathbf{p}_{-k,n} = (p_{1,n}, \ldots, p_{k-1,n}, p_{k+1,n}, \ldots, p_{K,n})^T$ denotes the power profile of all users except $k$ over subcarrier $n$, and

$$\mu_{k,n}(\mathbf{p}_{-k,n}) = \frac{\left|\mathbf{g}_{k,n}^H \mathbf{h}_{kk,n}\right|^2}{\|\mathbf{g}_{k,n}\|^2 \sigma^2 + \sum_{j=1,j\neq k}^{K} \left|\mathbf{g}_{k,n}^H \mathbf{h}_{kj,n}\right|^2 p_{j,n}}. \tag{4}$$

Using (3), the achievable rate (normalized to the subcarrier bandwidth, and thus measured in b/s/Hz) of the $k$-th user will be:

$$r_k(\mathbf{p}) = \frac{1}{N} \sum_{n=1}^{N} \log_2 \left(1 + \gamma_{k,n}\right) \tag{5}$$

where $\mathbf{p}_k = (p_{k,1}, \ldots, p_{k,N})$ denotes the power profile of user $k$ over all subcarriers $n = 1, \ldots, N$, and $\mathbf{p} = (\mathbf{p}_1, \ldots, \mathbf{p}_K) \in \mathbb{R}_+^{K \times N}$ is the corresponding power profile of all users (obviously, $p_{k,n} = 0$ if user $k$ is not transmitting over subcarrier $n$). To simplify notation, the argument of $\mu_{k,n}$ and $r_k$ will be suppressed in what follows.

### B. Problem Formulation

As mentioned in Section I, *energy-efficient* network design must take into account the energy consumption incurred by each UE. To that end, note that, in addition to the radiated powers $\mathbf{p}_k$ at the output of the radio-frequency front-end, each terminal $k$ also incurs circuit power consumption during transmission, mostly because of power dissipated at the UE signal amplifier [20, 22, 32]. Therefore, the overall power consumption $P_{T,k}$ of the $k$-th UE will be given by

$$P_{T,k} = p_{c,k} + P_k = p_{c,k} + \sum_{n=1}^{N} p_{k,n}, \tag{6}$$

where $P_k = \sum_{n=1}^{N} p_{k,n}$ is the transmitted power of user $k$ over the entire spectrum, while $p_{c,k}$ represents the average power consumed by the device electronics of the $k$-th UE (assumed for simplicity to be independent of the transmission state). Following [22, 33], the *energy efficiency* of the link can then be measured (in b/J/Hz) by the utility function

$$u_k(\mathbf{p}) = \frac{r_k}{P_{T,k}} = \frac{N^{-1} \sum_{n=1}^{N} \log_2 \left(1 + \mu_{k,n} p_{k,n}\right)}{p_{c,k} + \sum_{n=1}^{N} p_{k,n}} \tag{7}$$

where the dependence on the transmit power vectors of all other users is subsumed in the gains $\boldsymbol{\mu}_k = \{\mu_{k,n}\}_{n=1}^{N}$ of (4). Accordingly, in data-oriented wireless networks, QoS requirements take the form $r_k \geq \theta_k$, where $\theta_k$ is the minimum rate threshold required by user $k$.

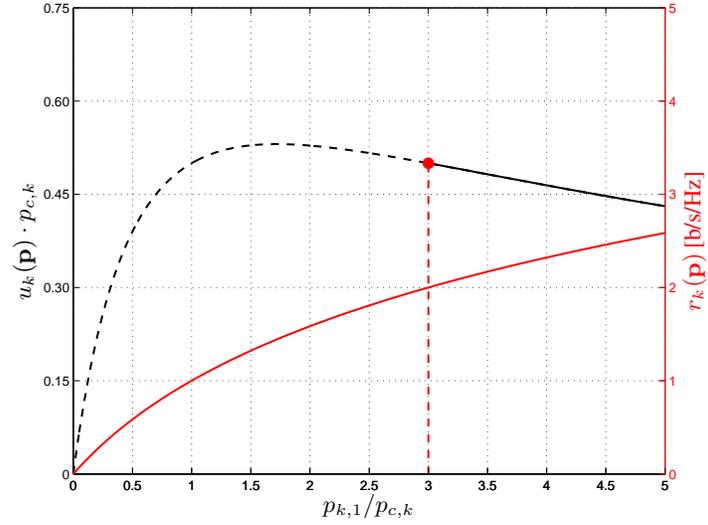

(a) $\mu_k \cdot p_{c,k} = 1$.

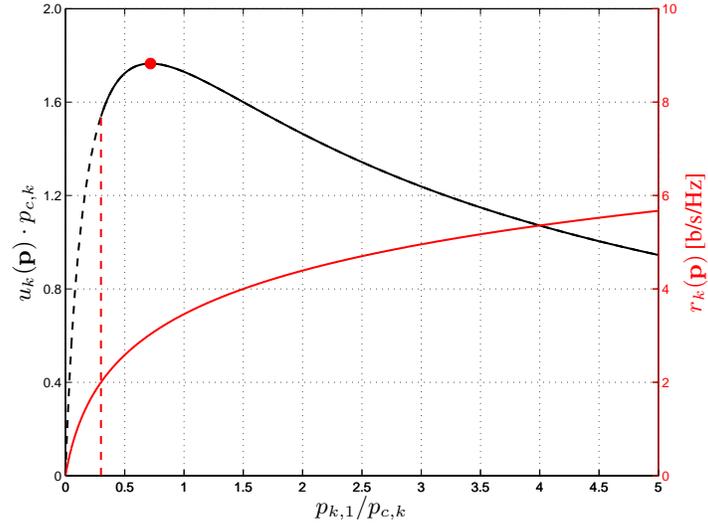

(b) $\mu_k \cdot p_{c,k} = 10$.

Fig. 1. Normalized utility as a function of the normalized transmit powers ($N = 1$, $\theta_k = 2$ b/s/Hz).

To summarize, the design of an energy-efficient resource allocation scheme which encompasses both subcarrier allocation and power control amounts to solving the following multi-agent, multi-objective optimization problem:

$$\text{maximize} \quad u_k(\mathbf{p}), \tag{8a}$$

$$\text{subject to} \quad N^{-1} \sum_{n=1}^{N} \log_2 \left( 1 + \mu_{k,n} p_{k,n} \right) \geq \theta_k, \tag{8b}$$

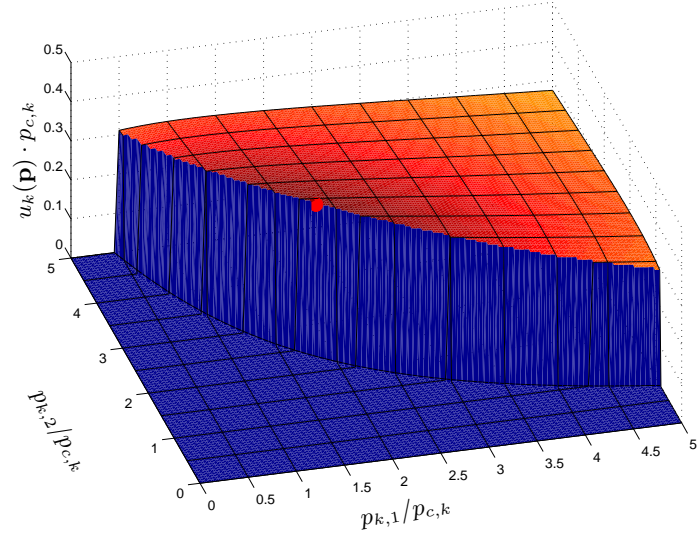

(a) $\boldsymbol{\mu}_k \cdot p_{c,k} = (1, 2)$.

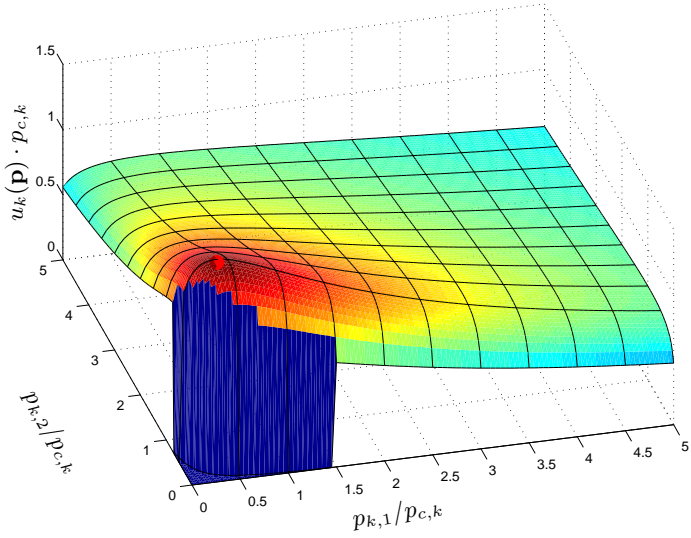

(b) $\boldsymbol{\mu}_k \cdot p_{c,k} = (10, 20)$.

Fig. 2. Normalized utility as a function of the normalized transmit powers ($N = 2$, $\theta_k = 2\,\text{b/s/Hz}$).

where $u_k(\mathbf{p})$ is the energy efficiency utility function (7) and (8b) represents the normalized rate requirement. Thus, unlike other OFDMA resource allocation problems (such as [34, 35]), subcarrier selection and power loading are tackled in a *joint* manner. Furthermore, inter- and intra-cell interference between UE transforms (8) into a game where each UE $k \in \mathcal{K}$ aims at unilaterally maximizing its individual link energy-efficiency via an optimal choice of power allocation vector $\mathbf{p}_k$ – and, in so doing, obviously affects the possible choices of all other UE in the network.

*Remark* 1. To visualize the impact of the rate constraints (8b) on the optimization problem (8), Figs. 1 and 2 depict the graph of the utility function (7) of user $k$ (normalized by $p_{c,k}$) as a function of the transmit powers $\mathbf{p}_k = \{p_{k,n}\}_{n=1}^N$ for a fixed interference power vector $\mathbf{p}_{-k}$ (and hence keeping $\{\mu_{k,n}(\mathbf{p}_{-k})\}_{n=1}^N$ fixed). For the sake of visualization, Fig. 1 depicts only $N = 1$ subcarrier. The dashed black line depicts the unconstrained utility (7), whereas the solid black line reports $u_k(\mathbf{p})$ for the values of $p_{k,1}$ such that (8b) holds, assuming $\theta_k = 2$ b/s/Hz (for convenience, also the rate $r_k$ is reported with red lines): $\mu_{k,1} = 1/p_{c,k}$ in Fig. 1(a), whereas $\mu_{k,1} = 10/p_{c,k}$ in Fig. 1(b). As can be seen, the power level that maximizes $u_k(\mathbf{p})$ (red dot) is on the left boundary of the feasible power set of Fig. 1(a): in this case, maximizing $u_k(\mathbf{p})$ corresponds to minimizing the power subject to rate constraints, e.g., as considered in [29]. In general however, the maximization of energy efficiency produces a different optimal point, as reported in Fig. 1(b) where the focal user can exploit better channel conditions experienced to increase its utility. This formulation is particularly appealing for next-generation wireless systems [27], as it captures the tradeoff between obtaining a satisfactory spectral efficiency and saving as much energy as possible [19, 22, 33]. This behavior is analogous to what can be observed in Fig. 2 where $N = 2$ and $\theta_k = 2$ b/s/Hz. When the channel conditions are not favorable (in Fig. 2(a), $\boldsymbol{\mu}_k \cdot p_{c,k} = (1, 2)$), the optimal power allocation $\mathbf{p}_k/p_{c,k} = (1.83, 2.33)$ lies on the contour of the (normalized) utility surface that guarantees $r_k(\mathbf{p}) \geq \theta_k$ (when $r_k(\mathbf{p}) < \theta_k$, we assume here $u_k(\mathbf{p}) = 0$ for the sake of graphical representation) – thus getting $r_k(\mathbf{p}) = \theta_k$. On the contrary, when the channel conditions are more favorable (in Fig. 2(b), $\boldsymbol{\mu}_k \cdot p_{c,k} = (10, 20)$), the utility is maximized by $\mathbf{p}_k/p_{c,k} = (0.37, 0.42)$, that yields $r_k(\mathbf{p}) = 2.74$ b/s/Hz $> \theta_k$.

*Remark* 2. It is easy to see that a particular set of constraints $\{\theta_k\}_{k=1}^K$ may affect the *feasibility* of the problem in the sense that there might not exist *any* power allocation $\mathbf{p} \in \mathbb{R}_+^{K \times N}$ that allows all constraints $\theta_k$ to be met *simultaneously* – essentially due to mutual interference in the network, which implies a dependence between the gains $\boldsymbol{\mu}_k \ \forall k$. Necessary and sufficient conditions that ensure the feasibility of the problem (8) in the single-carrier case $N = 1$ can be found in [21]. On the other hand, analogous conditions for the general case of $N > 1$ subcarriers are very difficult to obtain, and future investigations will focus on addressing this issue.

# III. Game-Theoretic Resource Allocation

## A. Game-theoretic formulation of the problem

As mentioned earlier, mutual interference in the network introduces interactions among the users that aim at optimizing their utilities (8). A natural framework for studying such strategic inter-user interactions is offered by the theory of non-cooperative games with continuous (and action-dependent) action sets. Thus, following Debreu [24] (see also [25]), we will formulate the problem as a non-cooperative game $\mathcal{G} \equiv \mathcal{G}(\mathcal{K}, \mathcal{P}, u)$ consisting of the following components:

a) The set of *players* of $\mathcal{G}$ is the set $\mathcal{K}$ of the network's UE.

b) A priori, each player can choose any transmit power vector in $\mathcal{P}_k^0 \equiv \mathbb{R}_+^N$. However, given a power profile $\mathbf{p}_{-k} \in \mathcal{P}_k^0 \equiv \prod_{\ell \neq k} \mathcal{P}_\ell^0$ of the opponents of player $k$, the *feasible action set* of player $k$ in the presence of the rate requirements (8b) is:

$$\mathcal{P}_k(\mathbf{p}_{-k}) = \left\{ \mathbf{p}_k \in \mathcal{P}_k^0 : r_k(\mathbf{p}) \geq \theta_k \right\}. \tag{9}$$

c) The *utility* $u_k(\mathbf{p}_k; \mathbf{p}_{-k})$ of player $k$ is given by (7).

In this framework, the most widely used solution concept is a generalization of the notion of Nash equilibrium [4], known as *Debreu equilibrium* (DE) [24] and sometimes also referred to as *generalized Nash equilibrium* (GNE) [25]. Formally:

**Definition 1.** A power profile $\mathbf{p}^\star$ is a *Debreu equilibrium* of the energy-efficiency game $\mathcal{G}$ if

$$\mathbf{p}_k^\star \in \mathcal{P}_k(\mathbf{p}_{-k}^\star) \quad \forall k \in \mathcal{K}, \tag{10a}$$

and

$$u_k(\mathbf{p}^\star) \geq u_k(\mathbf{p}_k; \mathbf{p}_{-k}^\star) \quad \forall \mathbf{p}_k \in \mathcal{P}_k(\mathbf{p}_{-k}^\star), \ k \in \mathcal{K}. \tag{10b}$$

The main difference between Debreu and Nash equilibria is that the latter notion posits that players can unilaterally deviate to *any* feasible action, irrespective of whether this action satisfies the (coupled) constraints imposed on a player's action set by the actions of other players in the game. Put differently, Nash-type deviations include any action that satisfies a player's individual, *uncoupled* constraints, even if so doing violates the player's *coupled* constraints. In the case at hand, this means that, at Nash equilibrium, users would be allowed to transmit at any power level, even if this violates the system's transmission rate requirements. On the other hand, these feasibility constraints are already ingrained in the DE concept: the only unilateral deviations considered in (10b) are those for which the rate constraints are satisfied.[2]

As such, Debreu equilibria are of particular interest in the context of distributed systems because they offer a stable solution of the game from which players (in this case, UE) have no incentive to deviate (and thus destabilize the system) if everyone else maintains their chosen power allocation profiles. Accordingly, in what follows, we investigate the existence and characterization of DE in the energy-efficient power allocation game $\mathcal{G}$, leaving the question of uniqueness and convergence to such states to Sections III-C and IV, respectively.

### B. Problem feasibility and equilibrium existence

Debreu's original analysis [24] provides a general equilibrium existence result under the following assumptions:

(D1) The players' feasible action sets $\mathcal{P}_k(\mathbf{p}_{-k})$ are nonempty, closed, convex, and contained in some compact set $\mathcal{C}_k$ for all $\mathbf{p}_{-k} \in \mathcal{P}_{-k} \equiv \prod_{\ell \neq k} \mathcal{P}_\ell$.

(D2) The sets $\mathcal{P}_k(\mathbf{p}_{-k})$ vary continuously with $\mathbf{p}_{-k}$ (in the sense that the graph of the set-valued correspondence $\mathbf{p}_{-k} \mapsto \mathcal{P}_k(\mathbf{p}_{-k})$ is closed).

(D3) Each user's payoff function $u_k(\mathbf{p}_k; \mathbf{p}_{-k})$ is quasi-concave in $\mathbf{p}_k$ for all $\mathbf{p}_{-k} \in \mathcal{P}_{-k}$.

In our setting, $r_k(\mathbf{p}_k; \mathbf{p}_{-k})$ in (5) is concave in $\mathbf{p}_k$ and unbounded from above, so $\mathcal{P}_k(\mathbf{p}_{-k})$ is convex and nonempty for all $\mathbf{p}_{-k} \in \mathcal{P}_k^0$. Moreover, $\mathcal{P}_k(\mathbf{p}_{-k})$ varies continuously with $\mathbf{p}_{-k}$ because the constraints (8b) are themselves continuous in $\mathbf{p}_{-k}$. Finally, it is easy to show that $u_k(\mathbf{p}_k; \mathbf{p}_{-k})$ is quasi-concave in $\mathbf{p}_k$: since $u_k(\mathbf{p}_k; \mathbf{p}_{-k}) \geq a$ if and only if

$$r_k(\mathbf{p}_k; \mathbf{p}_{-k}) - a \left( p_c + \sum\nolimits_{n=1}^{N} p_{k,n} \right) \geq 0, \tag{11}$$

and the set defined by this inequality is convex for every $\mathbf{p}_{-k} \in \mathcal{P}_{-k}$ (recall that $r_k$ is concave in $\mathbf{p}_k$), quasi-concavity of $u_k(\,\cdot\,, \mathbf{p}_{-k})$ follows.

However, even though the users' best response sets

$$\mathcal{P}_k^\star(\mathbf{p}_{-k}) \equiv \underset{\mathbf{p}_k \in \mathcal{P}_k(\mathbf{p}_{-k})}{\arg\max} \; u_k(\mathbf{p}_k; \mathbf{p}_{-k}) \tag{12}$$

---

[2]The difference between Nash and Debreu equilibria is highlighted further if each player's transmit power is also constrained by a peak value (see below for more details): in this case, each user's individual power constraints would have to be satisfied by Nash-type deviations (and, of course, Debreu-type deviations as well), but Nash-type deviations would not necessarily satisfy the users' coupled QoS constraints.

are nonempty, convex, closed and bounded for every $\mathbf{p}_{-k}$, they might (and typically do) run off to infinity – i.e. they are not *uniformly* bounded. To understand this, simply consider the case of two UE transmitting over a single channel: if one of the UE transmits at very high power, the other UE is forced to transmit at a commensurately high power in order to meet its rate requirement. This leads to a cascade of power increases that makes each UE's feasible action set $\mathcal{P}_k(\mathbf{p}_{-k})$ (and, hence, $\mathcal{P}_k^\star(\mathbf{p}_{-k})$ as well) escape to infinity as the other UE increases its individual power. Formally, this means that the UE's feasible action sets $\mathcal{P}_k(\mathbf{p}_{-k})$ are not contained in an enveloping bounded set $\mathcal{C}_k$. Thus, Debreu's equilibrium existence theorem [24] does not apply.

From a power control perspective, this is not surprising: as is well known [36], the problem (8) may fail to be feasible, i.e. there may be no power profile $\mathbf{p} = (\mathbf{p}_1, \ldots, \mathbf{p}_K)$ such that $\mathbf{p}_k \in \mathcal{P}_k(\mathbf{p}_{-k})$ for all $k$. Obviously, in this case, the energy-efficiency game $\mathcal{G}$ does not admit an equilibrium either. On the other hand, at a purely formal level, equilibrium existence and problem feasibility are restored if we assume that users can transmit with infinitely high power, i.e. each UE $k \in \mathcal{K}$ chooses its total transmit power from the compactified half-line $[0, +\infty]$. In this extended setup, there are two points where indeterminacies may arise: first, the utility of player $k$ is not well-defined if $p_{k,n} = +\infty$ for some $n$; second, the rate requirement (8b) of user $k$ is also ill-defined if $p_{\ell,n} = +\infty$ for some $\ell \neq k$. To address these problems, note first that the utility function (7) of player $k$ decreases to 0 when $p_{k,n} \to +\infty$ for some channel $n = 1, \ldots, N$, reflecting the fact that $\lim_{x \to +\infty} x^{-1} \log_2 x = 0$. Thus, by continuity, the utility of player $k$ for infinite transmit powers $p_{k,n}$ may be defined as:

$$u_k(\mathbf{p}) = 0 \quad \text{whenever } p_{k,n} = +\infty \text{ for some } n. \tag{13}$$

As for the rate requirements of user $k$, a simple exponentiation of (8b) for finite $p$ yields the equivalent expression:

$$\prod_{n=1}^{N} \left(1 + \mu_{k,n} p_{k,n}\right) \geq 2^{N\theta_k} \tag{14}$$

or, after substituting for $\mu_{k,n}$ and rearranging:

$$\prod_{n=1}^{N} \left( \|\mathbf{g}_{k,n}\|^2 \sigma^2 + \sum_{j=1}^{K} \left|\mathbf{g}_{k,n}^H \mathbf{h}_{kj,n}\right|^2 p_{j,n} \right) \geq$$
$$2^{N\theta_k} \prod_{n=1}^{N} \left( \|\mathbf{g}_{k,n}\|^2 \sigma^2 + \sum_{j \neq k} \left|\mathbf{g}_{k,n}^H \mathbf{h}_{kj,n}\right|^2 p_{j,n} \right). \tag{15}$$

Since both sides of (15) are well-defined for all $p_{j,n} \in [0, +\infty]$, (15) provides a reformulation of (8b) that remains meaningful even in the extended arithmetic of $[0, +\infty]$.

In this infinite-power framework, any power profile $\mathbf{p}^\star = (\mathbf{p}_1^\star, \ldots, \mathbf{p}_K^\star)$ with $\sum_{n=1}^N p_{k,n}^\star = +\infty$ for all $k \in \mathcal{K}$ is feasible with respect to (15). Furthermore, if player $k$ deviates *unilaterally* and starts transmitting with finite total power, its rate requirement (15) will be automatically violated and its utility equals $0$. Consequently, no player can gain a utility greater than $0$ by deviating from $\mathbf{p}^\star$. This shows that the resulting infinite-power game $\overline{\mathcal{G}}$ with utility functions and rate requirements extended as in (13) and (15) above always admits a DE – and trivially so. However, any such equilibrium is clearly unreasonable from a practical standpoint as it represents a cascade of power increases that escapes to infinity as players try to meet their power constraints.

In view of the above, we could consider an alternative formulation of $\mathcal{G}$ in which the users' *uncoupled* action sets (i.e. unadjusted for the actions of other users) are of the form

$$\boldsymbol{\mathcal{P}}_k^0 = \left\{ \mathbf{p}_k \in \mathbb{R}_+^N : 0 \leq p_{k,n} \leq \overline{p}_{k,n}, \sum_n p_{k,n} \leq \overline{P}_k \right\} \tag{16}$$

for given maximum per-subcarrier transmit power levels $\overline{p}_{k,n}$ and total power constraints $\overline{P}_k$. In this case however, a crucial arising problem is that the resulting system could be even *unilaterally infeasible* in the sense that the admissible action set $\boldsymbol{\mathcal{P}}_k(\mathbf{p}_{-k})$ of player $k$ may be empty for a wide range of transmit power profiles $\mathbf{p}_{-k}$ of the other users in the system. Put differently, in the presence of maximum power constraints (a case that will be discussed at the end of Section IV), any given user may not be able to even participate in the game (in stark contrast with the formulation (9) of $\mathcal{G}$), thus exacerbating the equilibrium existence problem.

Of course, given that actual wireless devices cannot transmit at arbitrarily high levels, it is still crucial to determine under which conditions the game $\mathcal{G}$ admits a realizable DE. Therefore, in what follows, we will focus on conditions and scenarios, which guarantee that:

1) The energy-efficiency game $\mathcal{G}$ admits a DE with *finite* transmit powers (Section III-C).

2) This equilibrium is unique (Section III-C).

3) Users converge to equilibrium by following an adaptive, distributed algorithm (Section IV).

### C. Equilibrium characterization and uniqueness

The goal of this section is to characterize the game's DE by exploiting the fact that they are the fixed points of a certain best-response mapping.

**Proposition 1.** *A transmit power profile* $\mathbf{p}^\star$ *is at Debreu equilibrium if and only if its components* $p_{k,n}^\star$ *satisfy:*

$$p_{k,n}^\star = \left( \frac{1}{\lambda_k^\star} - \frac{1}{\mu_{k,n}} \right)^+ \tag{17}$$

*where*

$$\lambda_k^\star = \min \left\{ \lambda_k, \overline{\lambda}_k \right\}. \tag{18}$$

*In the above,*

$$\lambda_k = \frac{W \left( \alpha_k \cdot e^{\beta_k - 1} \right)}{\alpha_k} \tag{19}$$

*is the water level of the water-filling (WF) operator* (17) *when the problem* (8) *is solved without the minimum-rate constraints* (8b) *(i.e. when* $\theta_k = 0$ *for all* $k \in \mathcal{K}$*),* $W(\cdot)$ *denotes the Lambert W function [31], while*

$$\alpha_k = |\mathcal{S}_k|^{-1} \left( p_{c,k} - \sum_{n \in \mathcal{S}_k} \mu_{k,n}^{-1} \right) \tag{20}$$

*and*

$$\beta_k = |\mathcal{S}_k|^{-1} \sum_{n \in \mathcal{S}_k} \ln \mu_{k,n} \tag{21}$$

*where* $\mathcal{S}_k = \left\{ n \in \mathcal{N} : \mu_{k,n} \geq \lambda_k \right\}$ *denotes the subset of active subcarriers when using the uncostrained energy-efficient formulation. Similarly:*

$$\overline{\lambda}_k = \left( 2^{-N\theta_k} \prod_{n \in \overline{\mathcal{S}}_k} \mu_{k,n} \right)^{1/|\overline{\mathcal{S}}|_k} \tag{22}$$

*is the water level of* (17) *when* all *minimum-rate constraints* (8b) *are met simultaneously with equality (i.e.* (8) *reduces to a power minimization problem with equality rate constraints* $r_k = \theta_k$*), and, as above,* $\overline{\mathcal{S}}_k = \left\{ n \in \mathcal{N} : \mu_{k,n} \geq \overline{\lambda}_k \right\}$ *denotes the subset of active subcarriers.*

*Proof:* The proof is given in Appendix A and relies on defining the best-response mapping and using fractional programming to characterize its fixed points. ∎

*Remark* 3. Proposition 1 does not provide a way to calculate the water levels $\lambda_k$ and $\overline{\lambda}_k$. For an iterative computational method, the reader is referred to Section IV.

Despite its convoluted appearance, Proposition 1 is of critical importance from both a theoretical and practical point of view. Indeed, it is the basic step to derive sufficient conditions ensuring the existence and uniqueness of the DE and also to develop a distributed and scalable power allocation algorithm that steers the network to a stable equilibrium state.

To that end, note that the equilibrium characterization of Proposition 1 may be vacuous if the game does not admit a DE to begin with – for instance, if the original power control problem is not feasible. On that account, we have:

**Proposition 2.** *The energy-efficiency game $\mathcal{G}$ admits a unique DE $\mathbf{p}^\star$ whenever $\forall k \in \mathcal{K}$:*

$$\sum_{\substack{j=1 \\ j \neq k}}^{K} \sum_{n=1}^{N} \omega_{kj,n}^2 \sup_{\boldsymbol{\mu}_k \in \boldsymbol{\Omega}_k} \left[ \frac{1}{\varsigma_k^\star} \sum_{n \in \mathcal{S}_k^\star} \omega_{kk,n}^{-2} \left( \xi_{k,n}^2 + \varsigma_k^\star - 2\xi_{k,n} \right) \right] < 1 \tag{23}$$

*where $\boldsymbol{\Omega}_k = \prod_{n=1}^{N} (0, \sigma^{-2}\omega_{kk,n}]$, $\varsigma_k^\star = |\mathcal{S}_k^\star|$,*

$$\omega_{kj,n} = \frac{\left| \mathbf{g}_{k,n}^H \mathbf{h}_{kj,n} \right|^2}{\left\| \mathbf{g}_{k,n} \right\|^2} \tag{24}$$

*and*

$$\mathcal{S}_k^\star = \begin{cases} \overline{\mathcal{S}}_k & \text{if } \lambda_k \geq \overline{\lambda}_k \\ \mathcal{S}_k & \text{if } \lambda_k < \overline{\lambda}_k \end{cases} \tag{25}$$

$$\xi_{k,n} = \begin{cases} \mu_{k,n} \overline{\lambda}_k^{-1} & \text{if } \overline{\lambda}_k \leq \lambda_k \text{ and } n \in \mathcal{S}_k^\star \\ \frac{\mu_{k,n} - \lambda_k}{\lambda_k (1 + \nu_k)} & \text{if } \overline{\lambda}_k > \lambda_k \text{ and } n \in \mathcal{S}_k^\star \\ 0 & \text{if } n \notin \mathcal{S}_k^\star \end{cases} \tag{26}$$

*with $\nu_k = -\ln \lambda_k + (\beta_k - 1)$.*

*Proof:* The main steps for the proof are given in Appendices B and C; for a more detailed version, the reader is referred to the online technical report [37]. ∎

*Remark* 4. Notice that these sufficient conditions are similar to the well-known conditions ensuring the uniqueness of a Nash equilibrium in the non-cooperative rate maximization game studied by [9] in the context of the interference channel. Intuitively, (23) means that if the interfering connections for a user are sufficiently far away and the resulting SINR is high enough, then the DE exists and is unique. However, these conditions include a non-trivial optimization step w.r.t. $\boldsymbol{\mu}_k$ that depends on the actual opponents' power $\mathbf{p}_{-k}$. Indeed, the variables of the problem impact the values of $\lambda_k^\star$, $\mathcal{S}_k^\star$ and all functions $\xi_{k,n}$, making the conditions rather difficult

---

**Algorithm 1** Iterative algorithm to solve problem (8).

---

**set** $t = 0$

**initialize** $\mathbf{p}_k[t] = \mathbf{0}_N$ for all users $k \in \mathcal{K}$

**repeat**

    **for** $k = 1$ to $K$ **do**

        {loop over the users}

        **receive** $\{\gamma_{k,n}[t]\}_{n=1}^{N}$ from the serving AP

        **compute** $\lambda_k$ using Algorithm 2 and $\overline{\lambda}_k$ using inverse water-filling

        **set** $\lambda_k^\star = \min\left\{\lambda_k, \overline{\lambda}_k\right\}$

        **for** $n = 1$ to $N$ **do**

            {loop over the carriers}

            **update** $p_{k,n}[t+1] = (1/\lambda_k^\star - p_{k,n}[t]/\gamma_{k,n}[t])^+$

        **end for**

    **end for**

    **update** $t = t + 1$

**until** $\mathbf{p}_k[t] = \mathbf{p}_k[t-1]$ for all $k \in \mathcal{K}$

---

to be exploited. To tackle this issue, the online technical report [37] provides a set of sufficient conditions that are simpler. This is achieved by observing that the upper-bound of the supremum term in (23) boils down to computing a function of the system parameters only. The downside is that these simple conditions are more stringent than (23). Nevertheless, it is worth pointing out that the users of the network are never required to compute these conditions: (23) is only meant as a safety feature to guard against catastrophic system instabilities, to be calculated by the network administrator based on expected network usage scenarios.

*Remark* 5. Since the conditions of Proposition 2 are only sufficient, DE might exist even in the case where (23) does not hold for some $k \in \mathcal{K}$. As a matter of fact, when (8) is feasible, the distributed algorithm that we present in Section IV is observed to converge to a DE in all the numerical simulations performed and for every network scenario considered.

## IV. DISTRIBUTED IMPLEMENTATION

To derive a practical procedure allowing UE to reach the DE of $\mathcal{G}$ in a distributed fashion (*without* any distinction between SUE and MUE), we start by focusing on a specific UE $k \in \mathcal{K}$ and assume that all other UE $j \neq k$ have already chosen their optimal transmit powers $\mathbf{p}_{-k} = \mathbf{p}_{-k}^{\star}$ (in a possibly asynchronous fashion). From (4), we then see that the gains $\mu_{k,n}(\mathbf{p}_{-k,n}^{\star})$ needed to implement (17) are simply

$$\mu_{k,n}(\mathbf{p}_{-k,n}^{\star}) = \frac{\gamma_{k,n}}{p_{k,n}} \tag{27}$$

for all $n \in \mathcal{N}$. This means that the only information that is not locally available at the $k$-th UE to compute the optimal powers $\{p_{k,n}^{*}\}$ is the set of SINRs $\{\gamma_{k,n}\}$ measured at the serving SCA of the $k$-th UE, and which can be sent with a modest feedback rate requirement on the return channel (a discussion on the impact of a limited feedback can be adapted to this specific scenario from [38]).

Based on the above considerations, we can derive an iterative and fully decentralized algorithm to be adopted by each UE $k$ at each time step $t$ to solve the fixed-point system of equations (17) with a low-complexity, scalable and adaptive procedure. The pseudocode for the whole network is summarized in Algorithm 1. Note that, in practice, each UE $k \in \mathcal{K}$ only needs to implement the steps for only one value in the user loop (i.e., its own index), so the algorithm is suitable for asynchronous implementation in dynamic network configurations where each UE only requires the SINRs to be fed back by the serving SCA, without any further information on the network.

For the sake of clarity, the algorithm to compute $\lambda_k$ for each UE $k \in \mathcal{K}$ as in (19) is reported in Algorithm 2, whereas $\overline{\lambda}_k$ can easily be computed using standard inverse water-filling (IWF) methods [26]. Note that, although (19) is derived analytically in closed form and can be computed directly, it is still appealing to use the iterative procedure outlined in Algorithm 2, which takes advantage of the Dinkelbach approach [39] based on Newton's method. The latter is known to converge superlinearly for convex nonlinear fractional programming problems [39], and leads to substantial computational savings compared to evaluating the Lambert $W$ function directly. Interestingly, the Dinkelbach algorithm can also be properly modified to address the computation of the IWF-based quantity $\overline{\lambda}_k$, thus saving the complexity required for sorting the coefficients $\{\mu_{k,n}\}_{n=1}^{N}$ in a descending order [40]. For the sake of brevity, Algorithm 2 makes use of some functions that are introduced in the proof of Proposition 1 (Appendix A). For future reference, throughout the simulations reported in Section V, the convergence tolerance is set to $\varepsilon = 10^{-5}$,

---

**Algorithm 2** Iterative algorithm to compute $\lambda_k$ as in (19).

---

**set** a tolerance $\varepsilon \ll 1$

{initialization of the Dinkelbach method:}

**repeat**

 **select** a random $\lambda_k \in \mathbb{R}$

 **for** $n = 1$ to $N$ **do**

  **set** $p_{k,n} = (1/\lambda_k - p_{k,n}[t]/\gamma_{k,n}[t])^+$

 **end for**

 **compute** $\varphi(\mathbf{p}_k)$ and $\chi(\mathbf{p}_k)$ using (31) (see Appendix A)

 **set** $\Phi(\lambda_k) = \varphi(\mathbf{p}_k) - \lambda_k \chi(\mathbf{p}_k)$

**until** $\Phi(\lambda_k) \geq 0$

{Dinkelbach method:}

**while** $\Phi(\lambda_k) \geq \varepsilon$ **do**

 **set** $\lambda_k = \varphi(\mathbf{p}_k)/\chi(\mathbf{p}_k)$

 **for** $n = 1$ to $N$ **do**

  **set** $p_{k,n} = (1/\lambda_k - p_{k,n}[t]/\gamma_{k,n}[t])^+$

 **end for**

 **update** $\varphi(\mathbf{p}_k)$ and $\chi(\mathbf{p}_k)$ using (31)

 **set** $\Phi(\lambda_k) = \varphi(\mathbf{p}_k) - \lambda_k \chi(\mathbf{p}_k)$

**end while**

---

and we check whether the end state of the algorithm is a DE by testing the characterization of Proposition 1.

**Proposition 3.** *The iterates of Algorithm 1 converge to Debreu equilibrium whenever* (23) *holds.*

 *Proof:* The convergence of Algorithm 1 to an equilibrium point follows from the contraction properties of the best-response mapping investigated in Section III-C. ∎

*Remark* 6. Although the contraction properties of the best-response mapping are contingent on the sufficient conditions of Proposition 2, Algorithm 1 is still seen to converge to a DE of $\mathcal{G}$, provided that the problem is feasible to begin with (see the next section for a numerical assessment via extensive numerical simulations).

*Remark* 7. In the theoretical analysis of Section III (as well as in Algorithm 1), we consider neither total maximum power constraints $\overline{P}_k$, such that, $P_k \leq \overline{P}_k$, nor per-subcarrier maximum power constraints $\overline{p}_{k,n}$, such that $p_{k,n} \leq \overline{p}_{k,n}$. Although power masks are usually required by wireless standards to meet out-of-band emission policies, the power limits $\{\overline{P}_k\}_{k \in \mathcal{K}}$ and $\{\overline{p}_{k,n}\}_{k \in \mathcal{K}, n \in \mathcal{N}}$ significantly impact the analytical characterization of the DE $\mathbf{p}^\star$. For the sake of theoretical correctness, they are thus not included in the present work and are left as a future direction of research. However, it is worth stressing that: *i*) Algorithm 1 can easily accommodate $\{\overline{P}_k\}_{k \in \mathcal{K}}$ and $\{\overline{p}_{k,n}\}_{k \in \mathcal{K}, n \in \mathcal{N}}$, by setting $\lambda_k^* = \max\left\{\min\left\{\lambda_k, \overline{\lambda}_k\right\}, \underline{\lambda}_k\right\}$, where $\underline{\lambda}_k$ is computed using direct WF [26] (by maximizing the rate $r_k(\mathbf{p})$ under the constraint $\sum_{n=1}^{N} p_{k,n} = \overline{P}_k$), and by setting

$$p_{k,n}[t+1] = \min\left\{\overline{p}_{k,n}, \left(1/\lambda_k^* - p_{k,n}[t]/\gamma_{k,n}[t]\right)^+\right\};\tag{28}$$

*ii*) reasonable values of $\{\overline{P}_k\}_{k \in \mathcal{K}}$ and $\{\overline{p}_{k,n}\}_{k \in \mathcal{K}, n \in \mathcal{N}}$ do not modify the optimal power allocation $\mathbf{p}^\star$ in practice. In the interest of providing a practical algorithm that can be used in real-world scenarios, our extensive simulations in Section V make use of the modified algorithm, in which we observe that the selected values for the power constraints are never active in practice, so the theoretical results of Section III remain valid.

## V. Numerical results

Numerical simulations are now used to assess the performance of the proposed algorithm under different operating conditions. To keep the complexity of the simulations tractable while considering a significantly loaded system, we focus on the scenario reported in Fig. 3, where a square-shaped macrocell with an area of $200 \times 200\,\mathrm{m}^2$ centered around its MBS accommodates $S$ randomly distributed small cells, each with a radius of $\rho_s = \rho_S = 20\,\mathrm{m}$. Throughout the simulations, unless otherwise specified, we adopt the parameters reported in Table I (see [20] and references therein), where, for simplicity, each SC is assumed to have the same number of antennas $M_S$ and to serve the same number of users $K_S$. Moreover, all UE are assumed to have the same non-radiative power consumption $p_{c,k} = p_c$, and the same power limits $\overline{P}_k = \overline{P}$ and $\overline{p}_{k,n} = \overline{p}$ are imposed for all subcarriers (see Remark 7). To include the effects of fading and shadowing, we use the path-loss model introduced in [41], using a 24-tap channel model to

Table I

General system parameters

| Parameter | Value | Parameter | Value |
|-----------|-------|-----------|-------|
| Bandwidth | $B = 11.2$ MHz | Carrier spacing | $\Delta f = 10.9375$ kHz |
| Carrier frequency | $f_c = 2.4$ GHz | Macro-cell area | $0.04$ km$^2$ |
| Total number of small cells | $S = 5$ | Small-cell radius | $\rho_S = 20$ m |
| Number of antennas (MBS, SCA) | $M_0 = 16, M_S = 4$ | Density of population | $1{,}000$ users/km$^2$ |
| Number of SUE per small cell | $K_S = 4$ | Number of MUE | $K_0 = 20$ |
| Number of subcarriers | $N = 96$ | Noise power | $B\sigma^2 = -103.3$ dBm |
| Non-radiative power | $p_c = 20$ dBm | Path-loss exponent | $\zeta = 3.5$ |
| Cut-off parameter | $d_{ref} = 35$ m | Average path-loss attenuation at $d_{ref}$ | $L_{ref} = -84.0$ dB |
| Maximum total power | $\overline{P} = 40$ dBm | Maximum per-subcarrier power | $\overline{p} = 30$ dBm |

reproduce multipath effects. We also assume perfect channel estimation at the receiver end and the use of maximum ratio combining (MRC) techniques, which amounts to setting $\mathbf{g}_{k,n} = \mathbf{h}_{kk,n}$ for all $k \in \mathcal{K}$ and $n \in \mathcal{N}$. The UE $k \in \mathcal{K}$ is then assigned to APs $s \in \mathcal{S}$ following the mapping:

$$\psi(k) = \begin{cases} s & \exists\, s > 0 \text{ s.t. } d_{k,s} \leq \rho_S \\ 0 & \text{otherwise} \end{cases} \tag{29}$$

where $d_{k,s}$ denotes the distance between UE $k$ and SCA $s$. Without loss of generality, we measure the performance for a specific user (say user 1) within either an SC or a macrocell, by averaging over all possible positions of the users, uniformly randomizing their minimum-rate constraints $\theta_k$ in $[0, 2]$ [b/s/Hz] for $k \neq 1$.

To evaluate the proposed algorithm in a practical setting, Fig. 3 reports a random realization of the network with the parameters described above, in which the following quantities have been reduced for the sake of graphical representation: $K_S = 3$, $K_0 = 6$, and $N = 12$, $\theta_k = 1.5$ b/s/Hz for SUE, and $\theta_k = 0.5$ b/s/Hz for the MUE. Using the distributed algorithm described in Section IV, after roughly 20 iterations we get the solution to (8), representing the users' power profile at the DE of $\mathcal{G}$, and reported in Fig. 4. Here, the first five subplots correspond to the powers allocated in the small cells (the $s$-th subplot depicts the powers allocated by the users in the $s$-th small cell, with colors matching the ones used in Fig. 3), whereas the last two subplots show the powers selected by the MUE labeled $\{16, 17, 18\}$ (in the sixth subplot) and $\{19, 20, 21\}$ (in the seventh subplot), respectively. As can be seen in Fig. 4, this method tends to allocate the subcarriers in an exclusive manner whenever the MAI across UE within the same small cell is too large (e.g., see the $4$th small cell, in which only $5$ subcarriers are shared by the $3$ users), and to share the same subcarrier when the MAI across users is at a tolerable level (which also includes the interference generated by SUE from neighboring cells and the MUE). On the right hand side, we report the achieved rates at the DE in b/s/Hz. As can be verified, all users achieve their minimum demands, while for users with particularly favorable channel conditions (in this case, users no. $1$, $11$, $19$, and $21$), it is convenient to increase their transmit power so as to obtain better performance in terms of EE. As we mentioned in Section II, we assume the channel to be weakly time-varying. Otherwise stated, we assume that the convergence of the proposed algorithm is achieved before significant channel variations, as is customarily assumed in all closed-loop resource allocation schemes. To support this, assume that the uplink and downlink slot durations are in the order of few milliseconds (which is reasonable for LTE/LTE-A standards [42]). In these circumstances, the average convergence time of the proposed solution turns out to be in the order of tens of milliseconds (since convergence is achieved after approximately $20$ iterations): such interval is sufficiently shorter than typical channel coherence times, especially when considering usual SC scenarios with pedestrian users.

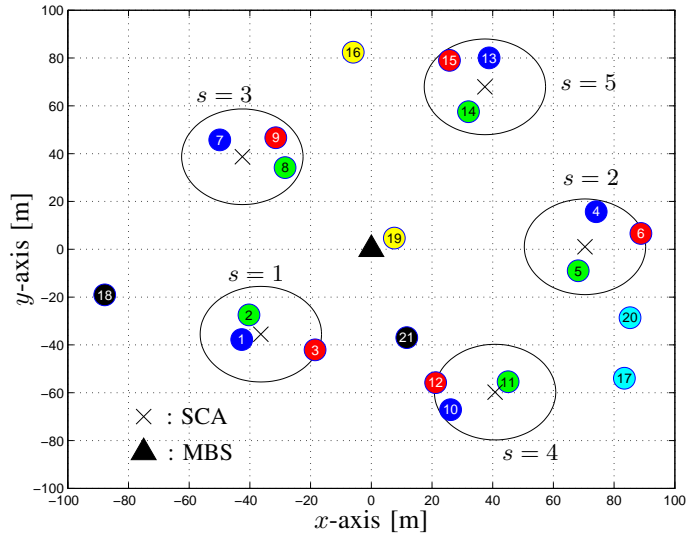

Fig. 3. Random realization of a network with $S = 5$ small cells, $K_S = 3$ SUE, and $K_0 = 6$ MUE, sharing $N = 12$ subcarriers.

To assess the robustness of the proposed solution to network perturbations, we depict in Fig. 5

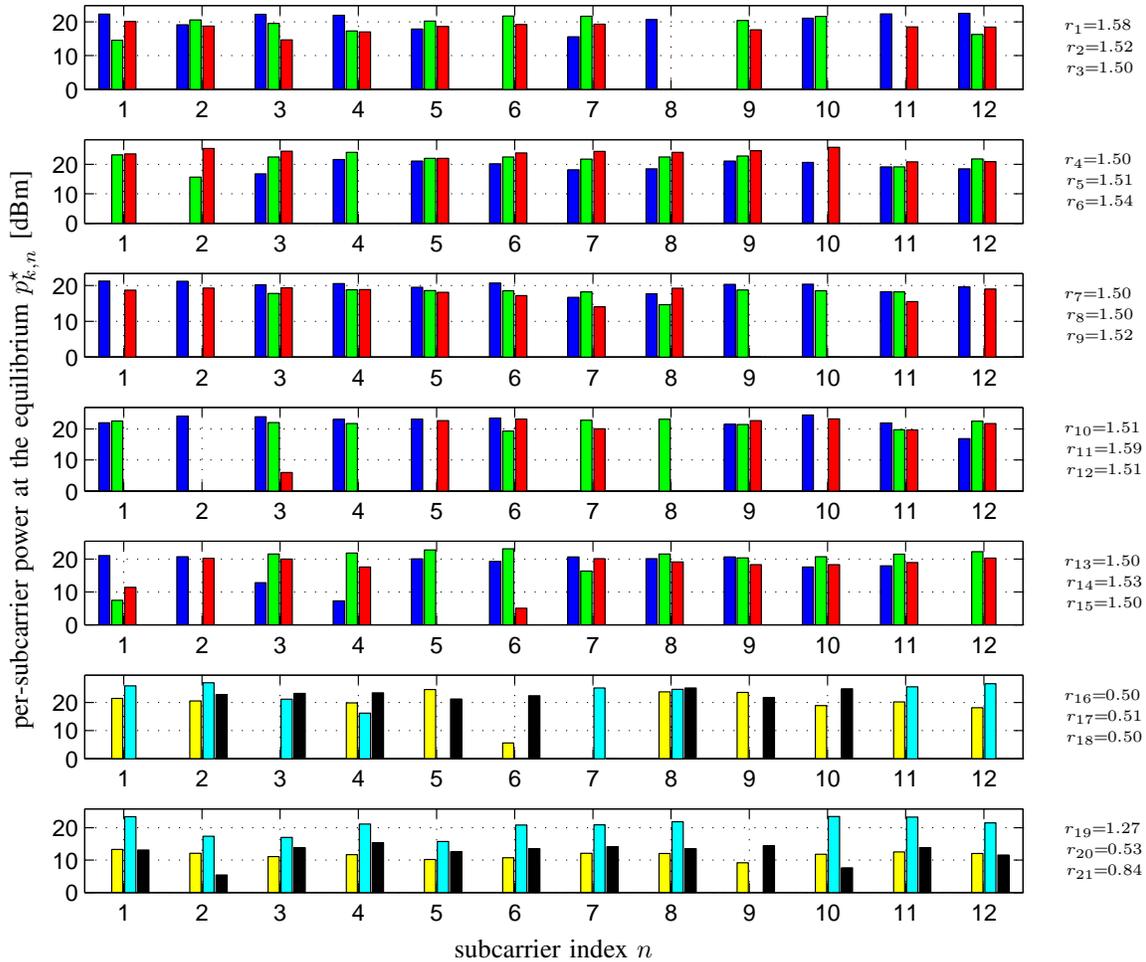

Fig. 4. Outcome of the resource allocation for the scenario of Fig. 3. The subcarriers are allocated exclusively when the MAI within the small cell is large. All users achieve their rate requirements. Users with favorable channels increase their powers to maximize their own utilities.

the total power consumption as a function of the iteration step for the network setting of Fig. 3 (lines are identified by UE labels, using the numbering adopted in Fig. 3). In particular, for the sake of clarity, since all other users show similar results, we only report the behavior of SUE in small cells $s = 1$ and $s = 4$, and the MUE 19 and 21, when, at $t = 25$, two cell-edge users (namely, users $\{3, 12\}$) simultaneously change their receiver association: both become served by the MBS, due to a variation in the received signal strength (with ensuing reduction of their data rate requirements to $0.5$ b/s/Hz, like all other MUE). As can be seen, the algorithm is very robust to network perturbations, and guarantees fast convergence for all users in the network to the new equilibrium point. In this particular example, each UE's power decrease is due to a lower interference generated by the "new" MUE – which, in turn, is a consequence of their lower target rates.

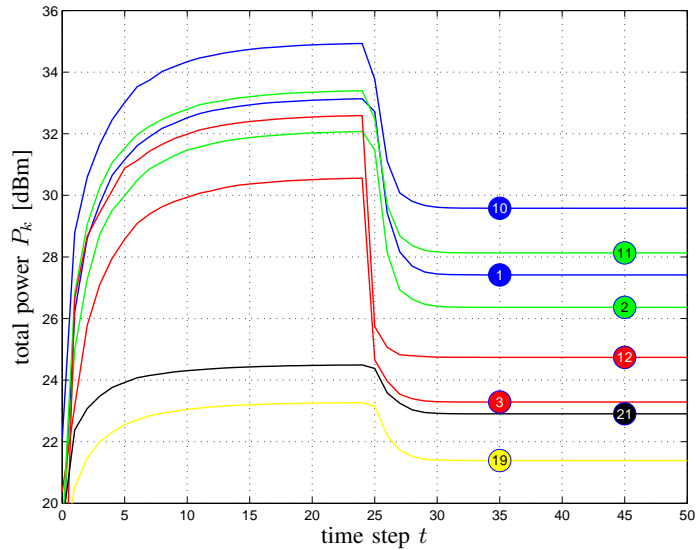

Fig. 5. UE total power consumption as a function of the iteration step. The power allocation fastly converge even in the presence of sudden changes in the network configuration, e.g., due to UE mobility or channel fluctuations.

To the best of our knowledge, there are no resource allocation algorithms that address the energy-efficient formulation (8) subject to the minimum-rate demands (8b). To evaluate the improvement in terms of EE of the proposed technique (red), we thus compare its performance with that achieved by an IWF-based solution (blue), in which all users aim at meeting $\theta_k$ with equality [29]. Fig. 6 reports the average utility achieved by averaging over all possible positions of a particular MUE (say user 1) as a function of a specific minimum rate $\theta_1$, using the parameters reported in Table I.[3] Interestingly, there exists a critical $\theta_1$ (in this case, $0.28$ b/s/Hz), for which the EE of IWF is higher than that achieved by the proposed formulation, mainly due to a weaker MAI caused by the IWF users, that transmit at lower powers than energy-efficient ones (not reported for the sake of brevity). However, IWF policies are *not* stable: if the network's UE adopt an IWF approach, then a UE that deviates from this criterion would *greatly increase its EE* (represented by the green line in Fig. 6). This situation is reminiscent of the well-known prisoner's dilemma [4] where there exist states with higher average utility, but which are obviously abandoned once a user deviates in order to maximize his individual benefits – and, hence, are inherently unstable in a non-cooperative, decentralized setting. In addition to this, the proposed approach shows two interesting properties compared to IWF: *i*) averaging over all network realizations and all minimum rates, Algorithm 1 achieves an average utility of $1.76$ Mb/J, which is larger than the IWF-based one, equal to $1.69$ Mb/J; and *ii*) it introduces fairness among the users, as its performance in terms of EE is weakly dependent on the QoS requirement $\theta_k$.

---

[3]Throughout all the simulations in the present and subsequent graphs, the selected parameters yield an occurrence of feasible scenarios, assessed a posteriori by letting each UE achieve their minimum-rate constraint (8b) with equality, larger than $99\%$. Once the scenario is checked to be feasible, the convergence of Algorithm 1 to a stationary point (a DE) occurs with probability 1.

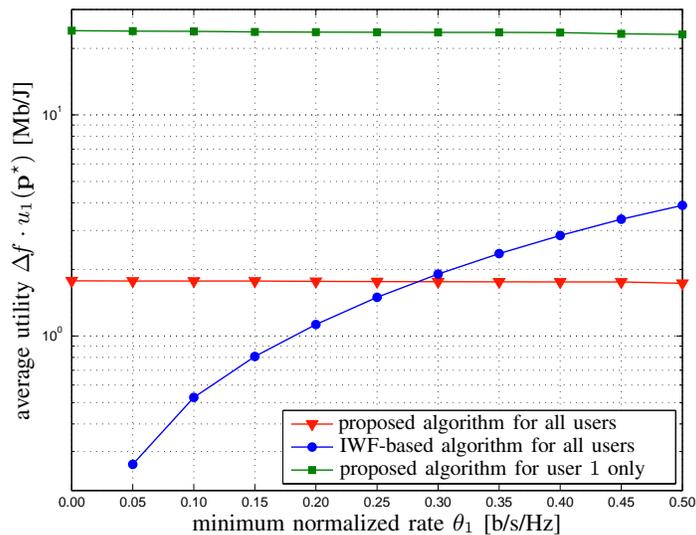

Fig. 6. Average utility at the equilibrium as a function of the minimum rate $\theta_1$. Compared to an IWF-based solution, the Debreu equilibrium may perform worse in terms of overall network utility. However, the IWF-based solution is not a stable operating point: user 1 has always an incentive to deviate and highly increase its own utility.

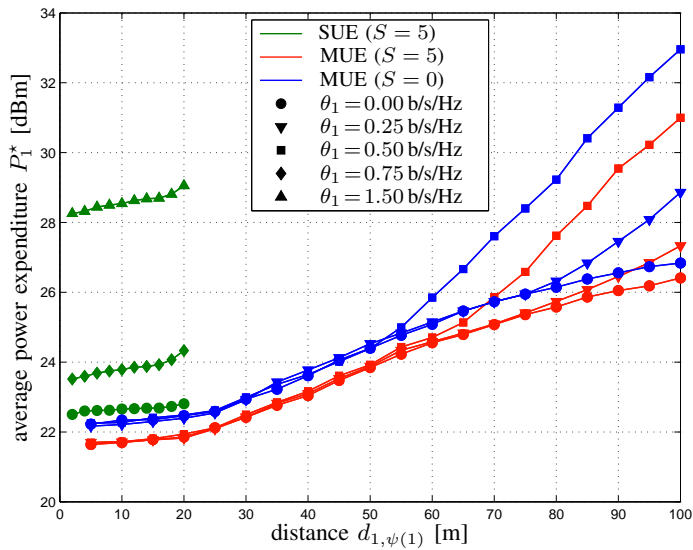

Fig. 7. Average transmit power at the equilibrium as a function of the distance from the receiver. The HetNet configuration ($S = 5$) significantly reduces the power consumption of the UE compared to the macro-cell classical scenario ($S = 0$) for any rate requirements.

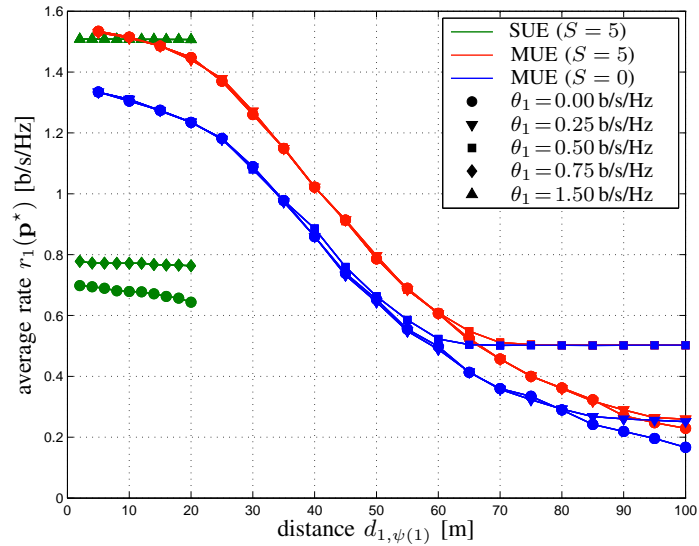

Fig. 8. Average rate at the equilibrium as a function of the distance from the receiver. The HetNet configuration ($S = 5$) significantly increases the rates of the UE compared to the macro-cell classical scenario ($S = 0$) for any rate requirements.

To measure the benefits of a HetNet configuration with respect to a classical macrocellular architecture ($S = 0$), Figs. 7 and 8 depict the average total transmit powers and the achievable rates at equilibrium in terms of the distance between the observed user and its receiver, averaged over $2,000$ independent feasible network realizations per marker. The green and red lines represent the performance in the case of $S = 5$ small cells, $K_S = 4$ SUE, and $K_0 = 20$ MUE, achieved by an SUE and an MUE, respectively, whereas blue lines show the performance obtained by an MUE in the case $S = 0$. We consider three different minimum demands for the SUE ($0$, $0.75$, and $1.5$ b/s/Hz, represented by circular, square, and upward-pointing arrowheads), and three different demands for the MUE (0, 0.25, and 0.5 b/s/Hz, represented by circular, downward-pointing arrowheads, and diamond markers respectively). As can be seen, the HetNet configuration introduces *significant gains in both the achievable rates and the power consumption* compared to the classical scenario: by averaging over all possible positions of SUE and MUE across the macrocell area, the MUE get $r_1(\mathbf{p}^\star) \approx 0.68$ b/s/Hz with a power consumption $P_1^\star \approx 27.5$ dBm (566 mW) when placing $\theta_1 = 0.5$ b/s/Hz,[4] compared to $r_1(\mathbf{p}^\star) \approx 0.63$ b/s/Hz with $P_1^\star \approx 29.1$ dBm (813 mW) for the same minimum demand in the case $S = 0$. The HetNet configuration is also *beneficial in terms of ASE*: using these parameters, we get on average slightly more than $600$ b/s/Hz/km², compared to $500$ b/s/Hz/km² for $S = 0$.

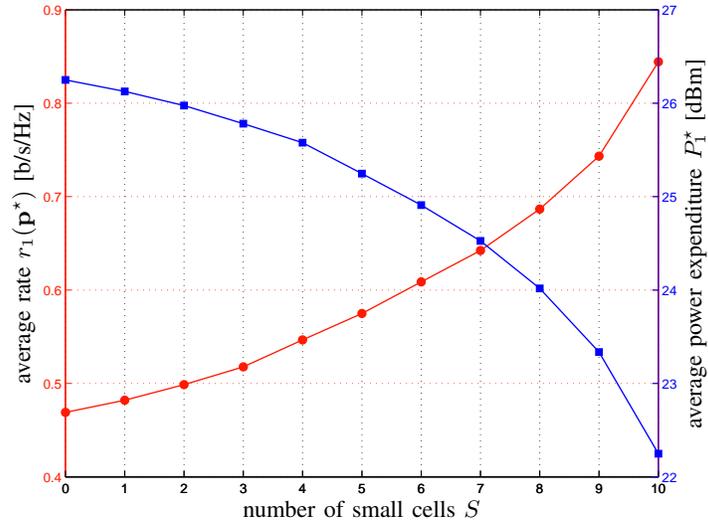

Fig. 9. Average rate at the equilibrium (left axis) and average power consumption (right axis) as functions of the number of small cells. Introducing more small cells increases the average rate and reduces the average power consumption in the network while guaranteeing the minimum rate requirements.

Introducing small cells has a negative impact in terms of the algorithm's convergence rate: here, on average $4.1$ iterations are required for the case $S = 5$, compared to $3.5$ for the case $S = 0$. This is due to decentralizing the resource allocation at each receiving station, thus slightly slowing the convergence of the algorithm. However, this provides a better MAI management ensured by SCAs, that allow SUE to obtain *higher rates with lower interfering powers* at the MBS. As can be seen, due to the path-loss model employed, which is roughly constant for distances within $d_{\text{ref}} > \rho_S$, the SUE performance is *independent of the distance from the SCA*. When SUE place $\theta_1 = 1.5$ b/s/Hz, the spectral efficiency is similar to that achieved by MUE located at comparable distance from the MBS (see Fig. 8), but at the cost of a larger power consumption (see Fig. 7): this is due to a better diversity at the receiver obtained by the MUE, since the MBS employes a larger number of antennas (16 versus 4). However, this does not hold true as the MUE distance increases: averaging over all positions, SUE obtain an average rate $r_1(\mathbf{p}^\star) \approx 1.51$ b/s/Hz (more than twice the MUE's one) using $P_1^\star \approx 28.6$ dBm (732 mW, slightly higher than MUE's one).

---

[4]Note that such minimum demand is about one order of magnitude larger than the one considered for cell-edge users in 4G networks, equal to 0.07 b/s/Hz [42] for a scarcely populated cell (at most 10 users).

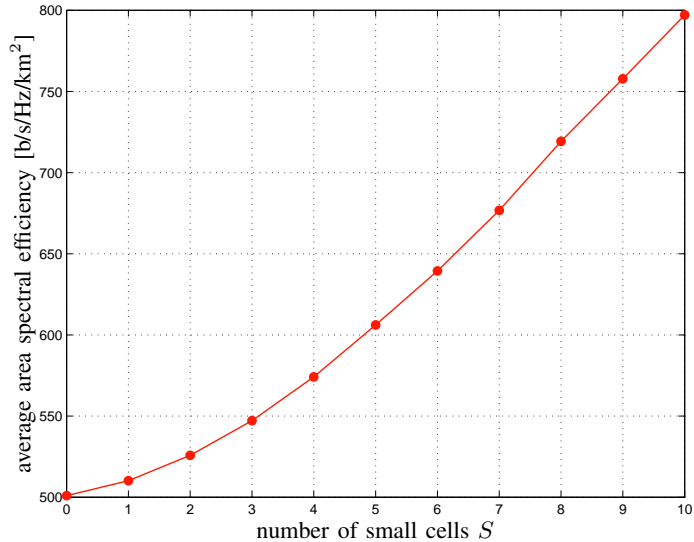

Fig. 10. Average area spectral efficiency as a function of the number of small cells. Introducing more small cells increases the average area spectral efficiency as well.

To emphasize the impact of small cells on the system performance, Figs. 9 and 10 compare the performance, averaged over $10^5$ independent network realizations, achieved by an MUE using $\theta_1 = 0.25$ b/s/Hz in the same network as before, populated by $K = 40$ users, as a function of the number of SCs $S$, each having $K_S = 4$ SUE, ranging from $S = 0$ (classical macrocell) to $S = 10$ (only SCs – in this case, the MUE of interest becomes an SUE). Fig. 9 depicts the achievable rate (red line, left axis) and the total power consumption (blue line, right axis), whereas Fig. 10 shows the ASE. As is apparent, introducing SCs in the system has a *significant benefit in terms of all performance indicators*. Of course, this comparison does not account for the additional complexity and drawbacks introduced by increasing $S$ (to mention a few, initial cost of network deployment and maintenance, and complexity of the system). However, although a suitable tradeoff needs to be sought, our analysis confirms that *network densification is one of the key technologies to meet 5G requirements* [27].

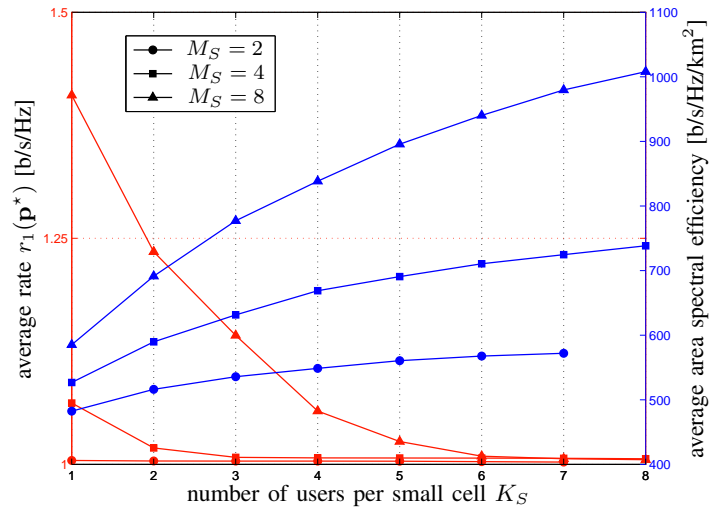

Fig. 11. Average rate (left axis) and average ASE (right axis) as functions of the number user per small cell. The average rate decreases with the number of users per small cell because of the MAI. However, the ASE is increasing with the the number of users per small cell. Moreover, increasing the number of receiving antennas at the SCA improves both, the average rate and average ASE.

To verify the scalability of the proposed solution, we also investigate the impact of the number of receiving antennas at the SCA $M_S$. In Fig. 11, we plot the spectral efficiency (red lines, left axis) and the ASE (blue lines, right axis) as a function of the number of users per small cell $K_S$. Circular, squared, and triangular markers represent the cases for $M_S = \{2, 4, 8\}$ antennas at the SCA. The ASE is averaged over all users $K = K_0 + S \cdot K_S$, whereas the achievable rate is computed for an SUE of interest using $\theta_1 = 1$ b/s/Hz, averaging over $10^5$ independent network realizations. As can be seen, increasing the number of antennas yields significant performance gains, thus representing a design parameter that can be exploited to boost the performance. Not only the spectral efficiency, as expected, benefits from increasing $M_S$ (as an example, we can move from $500$ b/s/Hz/km$^2$, achieved when using $2$ antennas, to $1,000$ b/s/Hz/km$^2$, by increasing the number of receiving antennas up to $8$, supporting $K = 60$ users), but *also does the EE*, confirming a recent result available in [32]: here, when $K_S = 7$, moving from $M_S = 2$ to $8$ yields more than a $5$-fold increase in the utility.

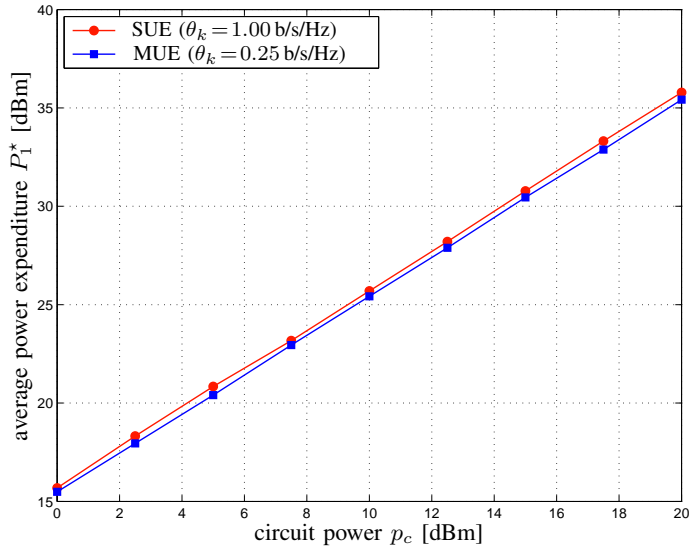

Fig. 12. Average power at the equilibrium as a function of the circuit power. The average power consumption scales linearly with the circuit power in the EE formulation.

Finally, to evaluate the impact of the circuit power $p_c$ on the EE of the system, we show in Fig. 12 the performance of the proposed algorithm as a function of $p_c$, averaged over $10^5$ independent network realizations, where the red line refers to an SUE using $\theta_k = 1$ b/s/Hz, and the blue line refers to an MUE using $\theta_k = 0.25$ b/s/Hz. For all selected non-radiative powers $p_c \in [0, 20]$ dBm, the hypothesis $p_c \gg \sigma^2$ holds, which is in line with the state of the art for radio-frequency and baseband transceiver modeling [20]. As can be seen, the total power consumption at the equilibrium $P_1(\mathbf{p}^\star)$ is directly proportional to $p_c$. Put differently, the energy-efficient equilibrium point is highly impacted by the non-radiative power, and the bit-per-Joule metric suggests the use a radiative power which is comparable with the non-radiative one. Interestingly, the (normalized) achievable rates at equilibrium (not reported for concision) do not depend on $p_c$ (1.1 and 0.6 b/s/Hz for SUE and MUE, respectively). This confirms a result which is well-known in the literature (e.g., see [22]): *EE increases as the circuit (non-radiative) power decreases*. Hence, reducing $p_c$, which is one of the main drivers in the device design further boosting the research in this field, can achieve a two-fold goal: not only is it expedient to reduce the constant power consumption (from an electronics point of view), but also it leads energy-aware terminals to reduce their radiative power when they aim at maximizing their bit-per-Joule performance (from an information-theoretic and resource-allocation perspective).

## VI. Conclusions and Perspectives

In this paper, we proposed a distributed power allocation scheme for energy-aware, non-cooperative wireless users with minimum-rate constraints in the uplink of a multicarrier heterogeneous network. The major challenge in this formulation is represented by the minimum-rate requirements that cast the problem into a non-cooperative game in the sense of Debreu, in which the actions sets of the players are coupled (and not independent as in the case of Nash-type games). We used fractional programming techniques to characterize the game's equilibrium states (when they exist) as the fixed points of a water-filling operator. To attain this equilibrium in a distributed fashion, we also proposed an adaptive, distributed algorithm based on an iterative water-filling best response process and we provided sufficient conditions for its convergence. The convergence and performance of the proposed solution were further assessed by numerical simulations: our results show that reducing the non-radiative power consumed by the user device electronics, offloading the macrocell traffic through small cells, and increasing the number of receive antennas, are critical to improve the performance of mobile terminals in terms of *both energy efficiency and spectral efficiency*. Using a realistic simulation setup, we showed that the proposed framework is able to achieve significantly high area spectral efficiencies (higher than $1,000$ b/s/Hz/km$^2$), peak and cell-edge spectral efficiencies (up to $6$ b/s/Hz and around $0.5$ b/s/Hz, respectively), and energy efficiencies (several Mb/J), while considering dense populations of users (around $1,000$ users/km$^2$), low power consumptions (at most a few Watts), a limited number of antennas (at most $8$ for the small-cell access points and $16$ for the macrocell base station), and simplified signal processing at the receiver (maximal ratio combining).

The system model adopted in this work encompasses a more general *multi-cellular* and *multi-tier* network, and the derived approach can be automatically adapted to such scenarios. Moreover, distinguishing features of the proposed distributed algorithm are its *scalability* and *flexibility*, which make it suitable for emerging 5G technologies [27], such as ultra-dense networks and massive MIMO.

Challenging open issues for further work include: *i)* assessing the feasibility of the problem given a particular network realization for the multicarrier case; *ii)* evaluating the impact of different receiver architectures (such as multiuser zero-forcing, and interference cancellation techniques) on the spectral and energy efficiency of the network; *iii)* accounting for highly time-varying scenarios in which users move around the network with high speeds.

## APPENDIX A

## PROOF OF PROPOSITION 1

First, note that (8) can be expressed in the language of fractional programming as:

$$\mathbf{p}_k^\star = \arg\max_{\mathbf{p}_k \in \mathcal{P}_k(\mathbf{p}_{-k})} \frac{\varphi(\mathbf{p}_k)}{\chi(\mathbf{p}_k)} \tag{30}$$

where $\mathcal{P}_k(\mathbf{p}_{-k})$ is defined as in (9), and

$$\varphi(\mathbf{p}_k) = \sum_{n=1}^{N} \ln(1 + \mu_{k,n} p_{k,n}) \text{ and } \chi(\mathbf{p}_k) = p_{c,k} + \sum_{n=1}^{N} p_{k,n}. \tag{31}$$

From [22, Sect. II.A] solving (30) is equivalent to finding the root of the following nonlinear function:

$$\Phi(\lambda_k) = \max_{\mathbf{p}_k \in \mathcal{P}_k(\mathbf{p}_{-k})} \varphi(\mathbf{p}_k) - \lambda_k \chi(\mathbf{p}_k) \tag{32}$$

where $\lambda_k \in \mathbb{R}$. To compute the solution of (30), let us first use (31), but without the constraint (8b), so that $\mathbf{p}_k \in \mathbb{R}_+^N$ (i.e., only nonnegative powers are considered). The stationarity condition, given by $\frac{\partial \varphi(\mathbf{p}_k)}{\partial p_{k,n}}|_{p_{k,n}=p_{k,n}^\star} - \lambda_k \frac{\partial \chi(\mathbf{p}_k)}{\partial p_{k,n}}|_{p_{k,n}=p_{k,n}^\star} = 0 \ \forall n$, using (31) becomes

$$\frac{\mu_{k,n}}{1 + \mu_{k,n} p_{k,n}^\star} - \lambda_k = 0 \qquad \forall n. \tag{33}$$

Hence, considering $p_{k,n}^\star \geq 0$, the optimal power allocation becomes the WF criterion (17), in which the water level $\lambda_k^\star$ is replaced by $\lambda_k$. By plugging (33) back into (32), we can finally compute the optimal power level $\lambda_k$:

$$-\ln \lambda_k + (\beta_k - 1) = \alpha_k \lambda_k \tag{34}$$

where the functions $\alpha_k$ and $\beta_k$ are defined as in (20) and (21), respectively. To provide a better insight on (34), let us define $\nu_k = -\ln \lambda_k + (\beta_k - 1)$, so that (34) can be rewritten as $\nu_k e^{\nu_k} = \alpha_k e^{\beta_k - 1}$. Using the Lambert function $W(\cdot)$ we can obtain the expression of $\lambda_k$ as in (19).

Introducing back the constraint (8b) simply places a lower bound on $\varphi(\mathbf{p}_k)$: $\varphi(\mathbf{p}_k) \geq \theta_k$. Following [22], this is equivalent to setting an upper bound $\overline{\lambda}_k$ on $\lambda_k$, that comes out of the IWF criterion that minimizes $\chi(\mathbf{p}_k)$ given $\varphi(\mathbf{p}_k) = \theta_k$, and is equal to (22). Hence, the solution to (8) is given by (17), with $\lambda_k^\star$ computed as in (18).

# Appendix B

## Proof of Proposition 2

There exists a unique DE $\mathbf{p}^\star$ if the best response map $\boldsymbol{\mathcal{B}}(\mathbf{p}) = [\boldsymbol{\mathcal{B}}_1(\mathbf{p}_{-1}), \ldots, \boldsymbol{\mathcal{B}}_K(\mathbf{p}_{-K})]$ with $\boldsymbol{\mathcal{B}}_k(\mathbf{p}_{-k}) = \arg\max_{\mathbf{p}_k \in \mathcal{P}_k(\mathbf{p}_{-k})} = u_k(\mathbf{p})$ is a contraction, , i.e., there exists some $\varepsilon \in [0,1)$ such that

$$\|\boldsymbol{\mathcal{B}}(\mathbf{p}_1) - \boldsymbol{\mathcal{B}}(\mathbf{p}_2)\| \le \varepsilon \|\mathbf{p}_1 - \mathbf{p}_2\| \qquad \forall \mathbf{p}_1, \mathbf{p}_2 \in \mathcal{P}, \tag{35}$$

where $\mathcal{P} = \prod_{k=1}^{K} \mathcal{P}_k$. The $n$th component of user $k$'s best response is given by $\mathcal{B}_{k,n}(\mathbf{p}^\star_{-k}) = [\boldsymbol{\mathcal{B}}_k(\mathbf{p}^\star_{-k})]_n = p^\star_{k,n}$ as in (17). We begin by rewriting $\mu_{k,n}(\mathbf{p}_{-k,n})$ in (4) as follows:

$$\mu_{k,n}(\mathbf{p}_{-k,n}) = \frac{\omega_{kk,n}}{\sigma^2 + I_{k,n}} \tag{36}$$

where $I_{k,n} = \sum_{j \ne k} \omega_{kj,n} p_{j,n}$, and the quantities $\omega_{kj,n}$ are defined in (24). Using [28, Theorem 4], the DE $\mathbf{p}^\star$ is unique if, for any UE $k$,

$$\left\| \frac{\partial \mathbf{I}_k}{\partial \mathbf{p}_{-k}} \right\| \cdot \sup_{\mathbf{I}_k \in \mathbb{R}^N} \left\| \frac{\partial \boldsymbol{\mathcal{B}}_k(\mathbf{p}_{-k})}{\partial \mathbf{I}_k} \right\| < 1 \tag{37}$$

with $\mathbf{I}_k = [I_{k,1}, \ldots, I_{k,N}]^T$. The first term of (37) is explicitly computed in [28, Eq. (19)], and it is equal to $\left\| \frac{\partial \mathbf{I}_k}{\partial \mathbf{p}_{-k}} \right\| = \sqrt{\sum_{j=1, j \ne k}^{K} \sum_{n=1}^{N} \omega_{kj,n}^2}$. As for the second term, we have:

$$\|\partial \boldsymbol{\mathcal{B}}_k(\mathbf{p}_{-k}) / \partial \mathbf{I}_k\| = \sqrt{\sum_{\ell=1}^{N} \sum_{n=1}^{N} \left| \partial p^\star_{k,n} / \partial I_{k,\ell} \right|^2}, \tag{38}$$

where the optimal (best-responding) transmit power levels $p^\star_{k,n}$ are:

$$p^\star_{k,n} = (1/\lambda^\star_k - 1/\mu_{k,n}) \mathbb{1}_{\{\mu_{k,n} > \lambda^\star_k\}}. \tag{39}$$

After some derivation steps, we obtain the norm of its partial derivative w.r.t. $I_{k,\ell}$ as follows:

$$\left| \frac{\partial p^\star_{k,n}}{\partial I_{k,\ell}} \right|^2 = \frac{\mathbb{1}_{\{\mu_{k,n} > \lambda^\star_k\}}}{\omega_{kk,\ell}^2 (\varsigma^\star_k)^2} \left[ \xi_{k,\ell}^2 + \left( (\varsigma^\star_k)^2 - 2\varsigma^\star_k \xi_{k,\ell} \right) \mathbb{1}_{\{n=\ell\}} \right] \tag{40}$$

where, for convenience, we denote by $\varsigma^\star_k = |\mathcal{S}^\star_k|$ and

$$\xi_{k,\ell} = -\varsigma^\star_k \mu^2_{k,\ell} \frac{\partial \left( 1/\lambda^\star_k \right)}{\partial \mu_{k,\ell}}. \tag{41}$$

Summing over $n = 1, \ldots, N$ then yields:

$$\left\| \frac{\partial \boldsymbol{\mathcal{B}}_k(\mathbf{p}_{-k})}{\partial \mathbf{I}_k} \right\| = \sqrt{\frac{1}{\varsigma^\star_k} \sum_{\ell \in \mathcal{S}^\star_k} \frac{1}{\omega^2_{kk,\ell}} \cdot \left( \xi_{k,\ell}^2 + \varsigma^\star_k - 2\xi_{k,\ell} \right)} \tag{42}$$

so it remains to show that the terms $\xi_{k,\ell}$ in (41) are equivalent to (26) in Proposition 2 (see Appendix C). As a final step in the proof, notice that the function to be optimized in (23) depends only on $\mu_{k,n}$ which is an invertible, bijective function of $I_{k,n} \geq 0$ (since it is a strictly decreasing function w.r.t. $I_{k,n}$). Therefore, we can take the supremum over $\mu_{k,n} \in (0, \omega_{kk,n}^2/\sigma^2], \forall n$ directly.

## Appendix C

In this section, we compute $\xi_{k,\ell}$ in two different cases depending on the relative order between $\lambda_k$ and $\overline{\lambda}_k$. Let us start from the minimum-rate WF criterion, in which UE $k$'s water level is computed using (18). In this case, if $\mu_{k,\ell} > \overline{\lambda}_k$ (i.e., if $\ell \in \overline{\mathcal{S}}_k$),[5] we have $\overline{\lambda}_k^{-1} = \left(2^{N\theta_k} \prod_{n \in \overline{\mathcal{S}}_k} \mu_{k,n}^{-1}\right)^{1/\overline{\varsigma}_k} = \left(2^{N\theta_k} \prod_{n \in \overline{\mathcal{S}}_k, n \neq \ell} \mu_{k,n}^{-1}\right)^{1/\overline{\varsigma}_k} \mu_{k,\ell}^{-1/\overline{\varsigma}_k}$, where $\overline{\varsigma}_k = |\overline{\mathcal{S}}_k|$. From this, we get $\frac{\partial (1/\overline{\lambda}_k)}{\partial \mu_{k,\ell}} = -\frac{1}{\overline{\varsigma}_k \mu_{k,\ell} \overline{\lambda}_k}$, and thus, using (41), we finally obtain $\xi_{k,\ell} = \mu_{k,\ell}/\overline{\lambda}_k$, corresponding to the first subcase of (26).

Let us now focus on the energy-efficient WF, in which each UE $k$'s water level is computed using (19). If $\mu_{k,\ell} > \lambda_k$, then:

$$
\begin{aligned}
\frac{\partial (1/\lambda_k)}{\partial \mu_{k,\ell}} &= \frac{1}{\lambda_k} \frac{\partial}{\partial \mu_{k,\ell}} \left[W\left(\alpha_k e^{\beta_k-1}\right) - (\beta_k - 1)\right] \\
&= \frac{1}{\lambda_k} \left[\frac{\partial W\left(\alpha_k e^{\beta_k-1}\right)}{\partial \mu_{k,\ell}} - \frac{\partial \beta_k}{\partial \mu_{k,\ell}}\right].
\end{aligned}
\tag{43}
$$

On one hand, using (20) and (21), we can compute the partial derivatives $\frac{\partial \alpha_k}{\partial \mu_{k,\ell}} = \frac{1}{\varsigma_k \mu_{k,\ell}^2}$ and $\frac{\partial \beta_k}{\partial \mu_{k,\ell}} = \frac{1}{\varsigma_k \mu_{k,\ell}}$, with $\varsigma_k = |\mathcal{S}_k|$. On the other hand, using the properties of the Lambert functions, we get

$$
\frac{\partial W\left(\alpha_k e^{\beta_k-1}\right)}{\partial \mu_{k,\ell}} = \frac{W\left(\alpha_k e^{\beta_k-1}\right) \cdot \frac{\partial}{\partial \mu_{k,\ell}}\left(\alpha_k e^{\beta_k-1}\right)}{\left(\alpha_k e^{\beta_k-1}\right)\left[1 + W\left(\alpha_k e^{\beta_k-1}\right)\right]}.
\tag{44}
$$

and hence:

$$
\frac{\partial (1/\lambda_k)}{\partial \mu_{k,\ell}} = \frac{W\left(\alpha_k e^{\beta_k-1}\right) - \alpha_k \mu_{k,\ell}}{\varsigma_k \mu_{k,\ell}^2 \lambda_k \alpha_k \left[1 + W\left(\alpha_k e^{\beta_k-1}\right)\right]}.
\tag{45}
$$

Noting that, by inverting (19), $W\left(\alpha_k e^{\beta_k-1}\right) = \beta_k - 1 - \ln \lambda_k$, and using simple mathematical steps, $\nu_k = -\ln \lambda_k + (\beta_k - 1)$ can be rewritten as $\nu_k = W\left(\alpha_k e^{\beta_k-1}\right) = \alpha_k \lambda_k$. Using (41), $\xi_{k,\ell}$ corresponds to the second subcase of (26).

---

[5] Note that we are interested in computing $\xi_{k,\ell}$ only for $\ell \in \overline{\mathcal{S}}_k$, as in all other cases $\xi_{k,\ell} = 0$.

# References

[1] "The 1000x data challenge," Qualcomm, Tech. Rep. [Online]. Available: http://www.qualcomm.com/1000x

[2] Green Touch Consortium, Tech. Rep. [Online]. Available: http://www.greentouch.org

[3] J. Hoydis, M. Kobayashi, and M. Debbah, "Green small-cell networks," *IEEE Veh. Technol. Mag.*, vol. 6, no. 1, pp. 37–43, 2011.

[4] D. Fudenberg and J. Tirole, *Game Theory*. Cambridge, MA: MIT Press, 1991.

[5] W. Yu, G. Ginis, and J. Cioffi, "Distributed multiuser power control for digital subscriber lines," *IEEE J. Sel. Areas Commun.*, vol. 20, no. 5, pp. 1105–1115, Jun. 2002.

[6] R. Cendrillon, J. Huang, M. Chiang, and M. Moonen, "Autonomous spectrum balancing for digital subscriber lines," *IEEE Trans. Signal Process.*, vol. 55, no. 8, pp. 4241–4257, Aug. 2007.

[7] F. Meshkati, H. V. Poor, and S. C. Schwartz, "An energy-efficient approach to power control and receiver design in wireless data networks," *IEEE Trans. Commun.*, vol. 53, no. 11, pp. 1885–1894, Nov. 2005.

[8] F. Meshkati, M. Chiang, H. V. Poor, and S. C. Schwartz, "A game-theoretic approach to energy-efficient power control in multicarrier CDMA systems," *IEEE J. Sel. Areas Commun.*, vol. 24, no. 6, pp. 1115–1129, Jun. 2006.

[9] G. Scutari, D. Palomar, and S. Barbarossa, "Optimal linear precoding strategies for wideband noncooperative systems based on game theory part I: Nash equilibria," *IEEE Trans. Signal Process.*, vol. 56, no. 3, pp. 1230–1249, Mar. 2008.

[10] ——, "Competitive design of multiuser MIMO systems based on game theory: A unified view," *IEEE J. Sel. Areas Commun.*, vol. 26, no. 7, pp. 1089–1103, Jul. 2008.

[11] P. Mertikopoulos, E. V. Belmega, and A. L. Moustakas, "Matrix exponential learning: Distributed optimization in MIMO systems," in *Proc. IEEE Intl. Symp. Information Theory*, Cambridge, MA, Jul. 2012, pp. 3028–3032.

[12] P. Coucheney, B. Gaujal, and P. Mertikopoulos, "Distributed optimization in multi-user MIMO systems with imperfect and delayed information," in *Proc. IEEE Intl. Symp. Information Theory*, Honolulu, HI, June-July 2014.

[13] Y. Shi, J. Wang, K. Letaief, and R. Mallik, "A game-theoretic approach for distributed power control in interference relay channels," *IEEE Trans. Wireless Commun.*, vol. 8, no. 6, pp. 3151–3161, Jun. 2009.

[14] S. Ren and M. van der Schaar, "Distributed power allocation in multi-user multi-channel cellular relay networks," *IEEE Trans. Wireless Commun.*, vol. 9, no. 6, pp. 1952–1964, Sep. 2010.

[15] D. T. Ngo, L. B. Le, T. Le-Ngoc, E. Hossain, and D. I. Kim, "Distributed interference management in two-tier CDMA femtocell networks," *IEEE Trans. Wireless Commun.*, vol. 11, no. 3, pp. 979–989, Mar. 2012.

[16] G. Scutari, D. Palomar, F. Facchinei, and J.-S. Pang, "Convex optimization, game theory, and variational inequality theory," *IEEE Signal Process. Mag.*, vol. 27, no. 3, pp. 35–49, Mar. 2010.

[17] J.-S. Pang, G. Scutari, D. Palomar, and F. Facchinei, "Design of cognitive radio systems under temperature-interference constraints: A variational inequality approach," *IEEE Trans. Signal Process.*, vol. 58, no. 6, pp. 3251–3271, Jun. 2010.

[18] E. V. Belmega and S. Lasaulce, "Energy-efficient precoding for multiple-antenna terminals," *IEEE Trans. Signal Processing*, vol. 59, no. 1, pp. 329–340, Jan. 2011.

[19] E. V. Belmega, S. Lasaulce, and M. Debbah, "A survey on energy-efficient communications," in *Proc. IEEE Intl. Symp. Personal, Indoor and Mobile Radio Communications Workshops (PIMRC)*, Istanbul, Turkey, Sep. 2010, pp. 289–294.

[20] E. Björnson, L. Sanguinetti, J. Hoydis, and M. Debbah, "Designing multi-user MIMO for energy efficiency: When is massive MIMO the answer?" in *Proc. IEEE Wireless Commun. and Networking Conf. (WCNC)*, Istanbul, Turkey, Apr. 2014.

[21] G. Bacci, E. V. Belmega, and L. Sanguinetti, "Distributed energy-efficient power optimization in cellular relay networks

with minimum rate constraints," in *Proc. IEEE Intl. Conf. Acoustics, Speech and Signal Process. (ICASSP)*, Florence, Italy, May 2014.

[22] C. Isheden, Z. Chong, E. Jorswieck, and G. Fettweis, "Framework for link-level energy efficiency optimization with informed transmitter," *IEEE Trans. Wireless Commun.*, vol. 11, no. 8, pp. 2946–2957, Aug. 2012.

[23] J. Nash, "Non-cooperative games," *Annals of Mathematics*, vol. 54, no. 2, pp. 286–295, 1951.

[24] G. Debreu, "A social equilibrium existence theorem," *Proc. Natl. Acad. Sci. USA*, vol. 38, no. 10, pp. 886–893, Oct. 1952.

[25] F. Facchinei and C. Kanzow, "Generalized Nash equilibrium problems," *Quarterly J. Operations Research*, vol. 5, no. 3, pp. 173–210, Sep. 2007.

[26] S. Boyd and L. Vandenberghe, Eds., *Convex Optimization*. Cambridge, UK: Cambridge Univ. Press, 2004.

[27] J. G. Andrews, S. Buzzi, W. Choi, S. Hanly, A. Lozano, A. C. K. Soong, and J. C. Zhang, "What will 5G be?" *IEEE J. Sel. Areas Commun.*, vol. 32, no. 6, pp. 1065–1082, Jun. 2014.

[28] G. Miao, N. Himayat, G. Li, and S. Talwar, "Distributed interference-aware energy-efficient power optimization," *IEEE Trans. Wireless Commun.*, vol. 10, no. 4, pp. 1323–1333, Apr. 2011.

[29] J.-S. Pang, G. Scutari, F. Facchinei, and C. Wang, "Distributed power allocation with rate constraints in gaussian parallel interference channels," *IEEE Trans. Information Theory*, vol. 54, no. 8, pp. 3471–3489, Aug. 2008.

[30] G. Bacci, E. V. Belmega, and L. Sanguinetti, "Distributed energy-efficient power and subcarrier allocation for OFDMA-based small cells," in *Proc. IEEE Intl. Conf. Commun. (ICC) Workshop on Small Cell and 5G Networks*, Sydney, Australia, Jun. 2014.

[31] R. Corless, G. Gonnet, D. Hare, D. Jeffrey, and D. Knuth, "On the Lambert W function," *Adv. Comp. Math.*, vol. 5, pp. 329–359, 1996.

[32] E. Björnson, L. Sanguinetti, J. Hoydis, and M. Debbah, "Optimal design of energy-efficient multi-user MIMO systems: Is massive MIMO the answer?" *IEEE Trans. Wireless Commun.*, 2015, to appear. [Online]. Available: http://arxiv.org/abs/1403.6150.

[33] G. Miao, N. Himayat, and G. Li, "Energy-efficient link adaptation in frequency-selective channels," *IEEE Trans. Commun.*, vol. 58, no. 2, pp. 545–554, Feb. 2010.

[34] C. Y. Wong, R. S. Cheng, K. B. Letaief, and R. D. Murch, "Multiuser OFDM with adaptive subcarrier, bit, and power allocation," *IEEE J. Select. Areas Commun.*, vol. 17, no. 10, pp. 1747–1758, Oct. 1999.

[35] H. Tabassum, Z. Dawy, and M. S. Alouini, "Sum rate maximization in the uplink of multi-cell OFDMA networks," in *Proc. Intl. Wireless Commun. and Mobile Computing Conf.*, Istanbul, Turkey, Jul. 2011.

[36] M. Chiang, P. Hande, T. Lan, and C. W. Tan, "Power control in wireless cellular networks," *Foundations and Trends in Networking*, vol. 2, no. 4, pp. 381–533, 2007.

[37] G. Bacci, E. V. Belmega, P. Mertikopoulos, and L. Sanguinetti, "Energy-aware competitive link adaptation in heterogeneous networks," Tech. Rep., Jul. 2014. [Online]. Available: http://www.iet.unipi.it/l.sanguinetti/hetnets.pdf

[38] G. Bacci, L. Sanguinetti, M. Luise, and H. V. Poor, "Energy-efficient power control for contention-based synchronization in OFDMA systems with discrete powers and limited feedback," *EURASIP J. Wireless Commun. and Networking (JWCN)*, vol. 2013, no. 1, Jul. 2013.

[39] W. Dinkelbach, "On nonlinear fractional programming," *Management Science*, vol. 13, no. 7, pp. 492–498, Mar. 1967.

[40] I. Stupia, L. Vandendorpe, L. Sanguinetti, and G. Bacci, "Distributed energy-efficient power optimization for relay-aided heterogeneous networks," in *Proc. Intl. Workshop Wireless Networks: Communication, Cooperation and Competition*, Hammamet, Tunisia, May 2014.

[41] G. Calcev, D. Chizhik, B. Göransson, S. Howard, H. Huang, A. Kogiantis, A. F. Molisch, A. L. Moustakas, D. Reed, and H. Xu, "A wideband spatial channel model for system-wide simulations," *IEEE Trans. Veh. Technol.*, vol. 56, no. 2, pp. 389–403, Mar. 2007.

[42] 3GPP Technical Specification Group, "LTE; Requirements for further advancements for Evolved Universal Terrestrial Radio Access (E-UTRA) (LTE-Advanced)," *Tech. Rep. 3GPP TR 36.913 v10.0.0*, Apr. 2011.